\documentstyle[aps,prd,epsf]{revtex}

\preprint{NUC-MINN-01/10-T}

\newcommand{\pslash}{\not{\! p}}

\newcommand{\be}{\begin{equation}}
\newcommand{\ee}{\end{equation}}
\newcommand{\ba}{\begin{eqnarray}}
\newcommand{\ea}{\end{eqnarray}}

\begin{document}
\draft

\title{Collective modes of color-flavor locked phase of dense QCD
at finite temperature}

\author{V. P. Gusynin}
\address{Bogolyubov Institute for Theoretical Physics,
        03143, Kiev, Ukraine \\
    and Department of Physics, Nagoya University,
        Nagoya 464-8602, Japan}

\author{I. A. Shovkovy\thanks{On leave of absence from
                    Bogolyubov Institute for Theoretical
                    Physics, 252143, Kiev, Ukraine.}}
\address{School of Physics and Astronomy, University of Minnesota,
        Minneapolis, MN 55455}

\date{\today}
\maketitle

\begin{abstract}
A detailed analysis of collective modes that couple to either vector or
axial color currents in color-flavor locked phase of color superconducting
dense quark matter at finite temperature is presented. Among the realm of
collective modes, including the plasmons and the Nambu-Goldstone bosons,
we also reveal the gapless Carlson-Goldman modes, resembling the scalar
Nambu-Golstone bosons. These latter exist only in a close vicinity of the
critical line. Their presence does not eliminate the Meissner effect,
proving that the system remains in the color broken phase. The finite
temperature properties of the plasmons and the Nambu-Goldstone bosons are
also studied. In addition to the ordinary plasmon, we also reveal a
``light" plasmon which has a narrow width and whose mass is of the order
of the superconducting gap. 
\end{abstract}




\section{Introduction}

In the last few years much progress has been achieved in understanding the
possible phases of quantum chromodynamics (QCD) at low temperatures and
high densities.  In particular it is expected that QCD at a sufficiently
large density of quark matter possesses color superconducting phases
\cite{BarFra,Bail}. It is widely believed that one of color
superconducting phases may exist even at moderate densities that
characterize the matter at the cores of compact stars (i.e., at densities
just a few times larger than the density of the ordinary nuclear matter).
The value of the superconducting order parameter is likely to be of order
$10$ to $100$ MeV \cite{W1,S1} (see also
Refs.~\cite{PR1,Son,us,SW2,PR2,H1,Br1,us2,Spt,PR-weak} for estimates using
the microscopic approach). These estimates of the order parameter are
encouraging to seek for observable signatures from a color superconducting
state inside compact stars \cite{stars}. While there are no many detailed
studies in this direction yet, a couple of issues have already been
addressed in Refs.~\cite{neutr-flux,Neutral,Interface}, and some of their
conclusions seem to be very promising.

At (asymptotically) large densities, the microscopic QCD becomes a weakly
interacting theory \cite{ColPer}. This allows one to use analytical
(although, non-perturbative) methods of quantum field theory for studying
dense quark matter. At finite chemical potential $\mu$, non-interacting
quarks should occupy all the states with momenta $|\vec{p}|<\mu$, thus,
forming the Fermi surface. In a more realistic (weakly) interacting model,
the attraction between quarks in the color-antitriplet channel leads to
the famous Cooper instability. The latter is removed by a rearrangement of
the ground state. As a result, in the case of the model with three quark
flavors, the original gauge symmetry $SU(3)_{c}$ and the global chiral
symmetry $SU(3)_{L} \times SU(3)_{R}$ break down to the global diagonal
$SU(3)_{c+L+R}$ subgroup \cite{CFL}. The corresponding phase is called the
color-flavor locked (CFL) phase. Out of total sixteen candidates for
Nambu-Goldstone (NG) bosons, eight are removed from the physical spectrum
by the Higgs mechanism, providing masses to eight gluons. The other eight
NG bosons show up as an octet (under the unbroken $SU(3)_{c+L+R}$) of
physical particles. In addition, the global baryon number symmetry as well
as the approximate $U(1)_{A}$ symmetry also get broken. As a result, an
extra NG boson and a pseudo-NG boson appear in the low energy spectrum.
These latter particles are both singlets under $SU(3)_{c+L+R}$.

The low energy effective action of the CFL phase was derived in
Refs.~\cite{CasGat,SonSt,HZB,Zar,Beane}, based on arguments of symmetry.
Moreover, all parameters of the action were calculated in the limit of the
asymptotically large chemical potential. The method of
Refs.~\cite{CasGat,SonSt,HZB,Zar,Beane} is based on matching vacuum
properties (such as vacuum energy and gluon screening) in the effective
and microscopic theories. While being very powerful for many purposes,
such a method is limited when it comes to determining the spectrum of
bound states other than NG bosons. The other approach used to study
(diquark) bound states is based on the Bethe-Salpeter (BS) equation
\cite{bs-cfl}. (A simplified study of the BS equation was also considered
in Ref.~\cite{Rho} and, for the case of two flavor dense QCD, in
Ref.~\cite{bs-short-long}.)

In this paper, we study the spectrum of collective excitations in the CFL
phase at finite temperature. To be more specific, we study only those
collective modes that couple to the (vector and axial) color currents. The
NG bosons and the massive plasmons are the examples of such modes.  As we
shall see, however, the plasmons and NG bosons do not exhaust all of the
collective modes. In addition, we also predict the existence of the
so-called Carlson-Goldman (CG) modes in the CFL phase of dense QCD (a
short outline of this prediction was given in Ref.~\cite{cg-short}). In
ordinary superconductors, the CG modes were experimentally discovered long
time ago \cite{CG}. These modes appear only in a close vicinity of the
critical line. Their quantum numbers are the same as those of the scalar
NG bosons that are removed from the physical spectrum by the Higgs
mechanism. Because of this, the appearance of the CG modes might look like
a revival of the NG bosons \cite{Tak97}.

This paper is organized as follows. In Sec.~\ref{model}, we describe our
model and introduce basic notations. Then, we present the general form of
the effective action of the gluon field and discuss its most important
properties in Sec.~\ref{eff-action}. In Sec.~\ref{pol-tensor}, we give the
explicit form of the finite temperature improved hard-dense loop
expression for the polarization tensor and reveal its internal structure.
Then, by making use of the polarization tensor, we derive the general form
of the current-current correlation function in Sec.~\ref{c-c-function}. In
the same section, we derive the gluon and the (order parameter) phase
field propagators and check that their poles appear at the same locations
as in the current-current correlation function. In Sec.~\ref{Spectra}, we
present a general formalism for studying collective modes, coupled to
vector (as well as axial) color currents. In Sec.~\ref{plasmon},
\ref{NG-boson} and \ref{CG-mode}, we analyze the properties of the
plasmons, the NG bosons and the gapless Carlson-Goldman modes,
respectively. Sec.~\ref{meissner-eff} deals with the Meissner effect in
the color superconducting phase. We derive the penetration depth of a
constant magnetic field, and show that the superconducting quark matter
corresponds to the so-called Pippard limit everywhere in the
superconducting part of the phase diagram except for a very small region
in the vicinity of the critical temperature. We summarize our results in
Sec.~\ref{conclusion}. Some useful notations and formulas are collected in
Appendices~\ref{AppZ} and \ref{AppA}. In Appendix~\ref{AppB}, we present
the general one-loop expression and the hard dense loop approximation of
polarization tensor. The key integrals, used in the definition of the
polarization tensor, are defined and approximately calculated in
Appendices~\ref{AppC} and \ref{AppD}.

\section{Model and notation}
\label{model}

As we mentioned in Introduction, the original $SU(3)_{c} \times SU(3)_{L}
\times SU(3)_{R}$ symmetry of massless QCD breaks down to the global
diagonal $SU(3)_{c+L+R}$ subgroup in the CFL ground state of dense
quark matter. The color condensate in the CFL phase is given by the vacuum
expectation value of the following diquark (antidiquark) field \cite{CFL}:
\be
\langle 0|  \left(\bar{\Psi}_{D}\right)^{a}_{i}
\gamma^{5} \left(\Psi_{D}^{C}\right)^{b}_{j} |0\rangle
=\kappa_{1} \delta^{a}_{i} \delta^{b}_{j}
+\kappa_{2} \delta^{a}_{j} \delta^{b}_{i}  ,
\label{order-par}
\ee
where $\Psi_{D}$ and $\Psi_{D}^{C}=C\bar{\Psi}^{T}_{D}$ are the Dirac
spinor and its charge conjugate spinor, and $C$ is a unitary matrix that
satisfies $C^{-1} \gamma_{\mu} C=-\gamma^{T}_{\mu} $ and $C=-C^{T}$. In
the last expression, we explicitly displayed the flavor ($i,j=1,2,3$) and
color ($a,b=1,2,3$) indices of the spinor fields.  The complex scalar
quantities $\kappa_{1}$ and $\kappa_{2}$ are determined by dynamics.
Throughout the paper, we follow the notation of Ref.~\cite{bs-cfl} as
close as possible.

It is convenient to introduce the color-flavor locked Weyl spinors
( octets and singlets under $SU(3)_{c+L}$ and $SU(3)_{c+R}$,
respectively) to replace the ordinary Dirac spinors,
\begin{mathletters}
\ba
\psi^{A}=\frac{1}{\sqrt{2}} {\cal P}_{+} (\Psi_{D})^{~i}_{a}
\left(\lambda^{A}\right)_{i}^{~a}, &\qquad&
\psi=\frac{1}{\sqrt{3}} {\cal P}_{+} (\Psi_{D})^{~i}_{a}
\delta_{i}^{~a}, \label{def-psi} \\
\tilde{\psi}^{A}=\frac{1}{\sqrt{2}} {\cal P}_{-}
(\Psi_{D}^{C})_{j}^{~b} \left(\lambda^{A}\right)_{b}^{~j}, &\qquad&
\tilde{\psi}=\frac{1}{\sqrt{3}} {\cal P}_{-}
(\Psi_{D}^{C})_{j}^{~b} \delta_{b}^{~j}, \label{def-psi-C} \\
\phi^{A}=\frac{1}{\sqrt{2}} {\cal P}_{-} (\Psi_{D})^{~i}_{a}
\left(\lambda^{A}\right)_{i}^{~a}, &\qquad&
\phi=\frac{1}{\sqrt{3}} {\cal P}_{-} (\Psi_{D})^{~i}_{a}
\delta_{i}^{~a}, \label{def-phi} \\
\tilde{\phi}^{A}=\frac{1}{\sqrt{2}} {\cal P}_{+}
(\Psi_{D}^{C})_{j}^{~b} \left(\lambda^{A}\right)_{b}^{~j}, &\qquad&
\tilde{\phi}=\frac{1}{\sqrt{3}} {\cal P}_{+}
(\Psi_{D}^{C})_{j}^{~b} \delta_{b}^{~j}, \label{def-phi-C}
\ea
\label{def-weyl}
\end{mathletters}
\noindent
where $A=1,\ldots,8$, and the sum over repeated indices
is understood. Tilde denotes the charge conjugate spinors. Also,
we use the conventional definition of the left- and right-handed
projectors, ${\cal P}_{\pm}=(1 \pm \gamma^5)/2$.

In the new notation, the non-zero order parameters $\kappa_{1}$ and
$\kappa_{2}$, as defined in Eq.~(\ref{order-par}), are related to
the following (singlet under the color-flavor locked residual symmetry)
vacuum expectation values:
\begin{mathletters}
\ba
\langle 0| \bar{\psi} \tilde{\psi} |0\rangle =
-\langle 0| \bar{\phi} \tilde{\phi} |0\rangle =
-\frac{1}{2} \left(3\kappa_{1} +\kappa_{2} \right) ,
\label{ord-par1} \\
\langle 0| \bar{\psi}^{A} \tilde{\psi}^{B} |0\rangle =
-\langle 0| \bar{\phi}^{A} \tilde{\phi}^{B} |0\rangle =
-\frac{1}{2} \delta^{AB} \kappa_{2} .
\label{ord-par2}
\ea
\label{ord-par}
\end{mathletters}
In general, the order parameters $\kappa_{1}$ and $\kappa_{2}$ are not
independent. The details of the pairing dynamics should define a specific
relation between the values of $\kappa_{1}$ and $\kappa_{2}$. The value of
their ratio, in its turn, defines a specific alignment of the order
parameter. The analysis of the SD equation in phenomenological
four-fermion models \cite{CFL}, as well as in the microscopic QCD theory
\cite{us2,Spt}, shows that the order parameter is dominated by the
antitriplet-antitriplet contribution ($\kappa_{1}=-\kappa_{2}$). While the
sextet-sextet admixture ($\kappa_{1}=\kappa_{2}$) is small, it is never
exactly zero. In fact, the latter does not break any additional symmetries
\cite{CFL}, and, for that reason alone, it cannot be excluded.  In our
analysis, we restrict ourselves to the case of a pure
antitriplet-antitriplet order parameter. We do this simply for the purpose
of convenience of presentation. The modification of our analysis to the
case with a nonzero sextet-sextet order parameter is straightforward. In
particular, it would only require to slightly change the ratio of the gaps
for the quasiparticle in the octet and singlet channels (see below).

The value of the order parameter in the CFL phase at zero temperature was
estimated in Refs.~\cite{CFL,us2,Spt}, using phenomenological four-fermion
as well as microscopic models [see Refs.~\cite{PR1,Son,us,SW2,PR2,H1,Br1},
dealing with the two flavor case]. While the exact value remains
uncertain, it can be as large as $100$ MeV. The temperature dependence of
the diquark order parameter is essentially the same as that in the
standard BCS theory of superconductivity \cite{PR-weak}. Notice that the
Meissner effect was neglected in the analysis of Ref.~\cite{PR-weak}.
However, the recent result of Ref.~\cite{R-glue} suggests that such an
approximation is justified.

For completeness of presentation, we note that the inverse form of the
relations in Eq.~(\ref{def-weyl}) reads
\begin{mathletters}
\ba
(\Psi_{D})_{a}^{~i} = \frac{1}{\sqrt{2}}
\left(\psi^{A} + \phi^{A}\right)
\left(\lambda^{A}\right)_{a}^{~i}
+\frac{1}{\sqrt{3}} \left(\psi + \phi\right)
\delta_{a}^{~i} \label{def-psi-D} \\
(\Psi_{D}^{C})_{j}^{~b} = \frac{1}{\sqrt{2}}
\left(\tilde{\psi}^{A} + \tilde{\phi}^{A}\right)
\left(\lambda^{A}\right)_{j}^{~b}
+\frac{1}{\sqrt{3}}
\left( \tilde{\psi}+\tilde{\phi}\right)
\delta_{j}^{~b} \label{def-psi-D-C}.
\ea
\end{mathletters}
Because the quarks in the color broken phase acquire a nonvanishing
Majorana mass, it is also useful to introduce the left-handed and
right-handed Majorana spinors,
\begin{mathletters}
\ba
\Psi = \psi+\tilde{\psi},  & \quad &
\Psi^{A} = \psi^{A}+\tilde{\psi}^{A}
\label{def-Maj-psi},\\
\Phi = \phi+\tilde{\phi},  & \quad &
\Phi^{A} = \phi^{A}+\tilde{\phi}^{A}
\label{def-Maj-phi},
\ea
\label{Maj-spinors}
\end{mathletters}
respectively. Then, in the CFL phase, the dynamically
generated Majorana
mass contribution takes a simple diagonal form,
\ba
{\cal L}_{\Delta} &=&
\frac{1}{2}\bar{\Psi} \left(
2\Delta {\cal P}_{-} + 2 \tilde{\Delta} {\cal P}_{+}
\right) \Psi
- \frac{1}{2}\bar{\Psi}^{A} \left(
\Delta  {\cal P}_{-} + \tilde{\Delta} {\cal P}_{+}
\right) \Psi^{A} \nonumber \\
&+&\frac{1}{2}\bar{\Phi} \left(
2\Delta {\cal P}_{+} + 2\tilde{\Delta} {\cal P}_{-}
\right)\Phi
- \frac{1}{2} \bar{\Phi}^{A} \left(
\Delta  {\cal P}_{+} + \tilde{\Delta} {\cal P}_{-}
\right)\Phi^{A}.
\label{L-mass}
\ea
Here $\Delta = \Delta^{+}_{T} \Lambda^{+}_{p} + \Delta^{-}_{T}
\Lambda^{-}_{p}$ and $\tilde{\Delta}=\gamma^0 \Delta^{\dagger}\gamma^0$,
and the quark ``on-shell"  projectors are
\ba
\Lambda_{p}^{\pm }=\frac{1}{2}
\left(1\pm \frac{\vec{\alpha} \cdot \vec{p}}{|\vec{p}|}\right),
\quad \vec{\alpha} =\gamma^{0} \vec{\gamma} .
\ea
Notice that the absolute value of the Majorana mass in Eq.~(\ref{L-mass})
in the singlet channel is twice as large as the that in the octet channel.
Also, the signs of those mass terms are opposite. Such a choice [see
Eq.~(\ref{ord-par})] corresponds to the assumption of a pure
antitriplet-antitriplet color-flavor locked order parameter
($\kappa_{1}=-\kappa_{2}$).

By making use of the Majorana spinors in Eq.~(\ref{Maj-spinors}), the QCD
action reads 
\ba 
{\cal L}_{QCD} &=& \frac{1}{2}\bar{\Psi} \left( \pslash
+\mu \gamma^0 \gamma^5 \right) \Psi 
+ \frac{1}{2}\bar{\Psi}^{A} \left(\pslash +\mu \gamma^0 \gamma^5 \right) 
\Psi^{A} 
+\frac{1}{2}\bar{\Phi} \left( \pslash -\mu \gamma^0 \gamma^5 \right) \Phi
+\frac{1}{2}\bar{\Phi}^{A} \left( \pslash -\mu \gamma^0 \gamma^5 \right)
\Phi^{A}    \nonumber \\ 
&& -\frac{1}{4}F_{\mu\nu}F^{\mu\nu} +\frac{g}{4}
\bar{\Psi}^{B} A^{A}_{\mu} \gamma^{\mu} \left(d^{ABC} \gamma^{5}
-if^{ABC}\right) \Psi^{C} +\frac{g}{2\sqrt{6}} A^{A}_{\mu}
\left(\bar{\Psi} \gamma^{\mu} \gamma^{5} \Psi^{A}+\bar{\Psi}^{A}
\gamma^{\mu} \gamma^{5} \Psi\right) \nonumber \\ 
&& -\frac{g}{4} \bar{\Phi}^{B} A^{A}_{\mu} \gamma^{\mu} 
    \left(d^{ABC} \gamma^{5} +if^{ABC}\right) \Phi^{C} 
-\frac{g}{2\sqrt{6}} A^{A}_{\mu} \left(
 \bar{\Phi} \gamma^{\mu} \gamma^{5} \Phi^{A}
+\bar{\Phi}^{A} \gamma^{\mu} \gamma^{5} \Phi\right). 
\label{L-QCD} 
\ea 
Here the structure constants $d^{ABC}$ and $f^{ABC}$ are defined by the
anticommutation and commutation relations of the Gell-Mann matrices,
\ba 
\left\{ \lambda^{A},
\lambda^{B} \right\} &=& \frac{4}{3} \delta^{AB} + 2 d^{ABC} \lambda^{C} ,
\\ \left[ \lambda^{A}, \lambda^{B} \right] &=& 2i f^{ABC} \lambda^{C} .
\ea 
By making use of the microscopic action in Eq.~(\ref{L-QCD}), we could 
derive the well known Schwinger-Dyson equation for the quark propagator in
the HDL improved, ladder approximation \cite{us2}. Its solution reveals
a dynamical generation of the Majorana quark masses. Such masses could be
incorporated in the low energy quark action by the contribution 
${\cal L}_{\Delta}$, presented in Eq.~(\ref{L-mass}). 

In this paper, we are primarily interested in studying the properties of
the collective modes at the non-perturbative vacuum in which the quarks
have the Majorana masses. A specific mechanism of the mass generation is
irrelevant for us here. Therefore, we could use the Hartree-Fock approach
in our analysis. In particular, we add and subtract the mass term ${\cal
L}_{\Delta}$ to the microscopic action in Eq.~(\ref{L-QCD}). Then, we
treat one of such terms as a part of the free quark propagator, while the
other one as an interaction term. In this approach, a special care should
be taken to preserve the gauge invariance of the model. This is because
the Majorana mass term in Eq.~(\ref{L-mass}) acquires phase factors under
a general gauge transformation, and thus, it is not gauge invariant. To
avoid the difficulty, it is convenient to consider the phases of the order
parameter $\Delta$ as quantum fields restoring the gauge invariance of the
model.

By taking into account the structure of the dynamically generated mass
terms in Eq.~(\ref{L-mass}), we arrive at the following quark propagators
in the CFL phase, 
\ba 
S_{1}(p) &=& i\frac{\gamma^0 (p_0+\epsilon_{p}^{+})-2\Delta^{+}_{T} }
{p_0^2-(\epsilon_{p}^{+})^2- 4|\Delta^{+}_{T}|^2} 
\Lambda_{p}^{-} {\cal P}_{+} 
+i\frac{\gamma^0 (p_0-\epsilon_{p}^{+})-2(\Delta^{+}_{T})^{*}}
{p_0^2-(\epsilon_{p}^{+})^2-4|\Delta^{+}_{T}|^2} 
\Lambda_{p}^{+} {\cal P}_{-} \nonumber \\ 
&+& i\frac{\gamma^0 (p_0-\epsilon_{p}^{-})-2\Delta^{-}_{T}}
{p_0^2-(\epsilon_{p}^{-})^2-4|\Delta^{-}_{T}|^2} 
\Lambda_{p}^{+} {\cal P}_{+} 
+i\frac{\gamma^0 (p_0+\epsilon_{p}^{-})-2(\Delta^{-}_{T})^{*}}
{p_0^2-(\epsilon_{p}^{-})^2-4|\Delta^{-}_{T}|^2} 
\Lambda_{p}^{-} {\cal P}_{-} , 
\label{S-1} \\ 
S_{8}^{AB}(p) \equiv \delta^{AB} S_{8}(p) &=& 
i \delta^{AB}\left( \frac{\gamma^0 (p_0+\epsilon_{p}^{+})+\Delta^{+}_{T} }
{p_0^2-(\epsilon_{p}^{+})^2-|\Delta^{+}_{T} |^2} 
\Lambda_{p}^{-} {\cal P}_{+} 
+\frac{\gamma^0 (p_0-\epsilon_{p}^{+})+(\Delta^{+}_{T})^{*}}
{p_0^2-(\epsilon_{p}^{+})^2-|\Delta^{+}_{T}|^2} 
\Lambda_{p}^{+} {\cal P}_{-} \right. \nonumber \\ 
&+&\left. \frac{\gamma^0 (p_0-\epsilon_{p}^{-})+\Delta^{-}_{T} }
{p_0^2-(\epsilon_{p}^{-})^2-|\Delta^{-}_{T}|^2} 
\Lambda_{p}^{+} {\cal P}_{+} 
+\frac{\gamma^0 (p_0+\epsilon_{p}^{-})+(\Delta^{-}_{T})^{*}}
{p_0^2-(\epsilon_{p}^{-})^2-|\Delta^{-}_{T}|^2} 
\Lambda_{p}^{-} {\cal P}_{-} \right) , 
\label{S-2} 
\ea 
for the left-handed fields. Similar expressions could also be written for
the right-handed fields (the latter are obtained from the above
expressions by the formal exchange of ${\cal P}_{-}$ and ${\cal P}_{+}$).
In the last two equations, we used the following notation:
$\epsilon_{p}^{\pm}=|\vec{p}|\pm\mu$. As is clear from the structure of
the quark propagators above, the value of the gap in the one-particle
quark spectrum is determined by $|\Delta^{-}_{T}|$. The so called antigap
$|\Delta^{+}_{T}|$, on the other hand, has an obscure physical meaning
because it can only be detected in the high energy ($q_{0} \agt 2 \mu$)
antiparticle excitations. In what follows we shall use the shorthand
notation $|\Delta_{T}| \equiv |\Delta^{-}_{T}|$ that should not create any
confusion.

\section{Effective action of gluons}
\label{eff-action}

In this paper, we are mostly interested in collective modes, coupled to
the (vector and axial) color currents. In the pursuit of a model
independent treatment, we find it advantageous to start our consideration
with the effective action, obtained by integrating out the quark degrees
of freedom. Such an action could be easily derived in a wide class of
four-fermion models of dense quark matter (in the weakly coupled regime).
As one could check, the effective Lagrangian density for the gluon field
should have the following general structure:
\ba
{\cal L}_{g,\phi}&=& -\frac{1}{2} A^{A,\mu}_{-q}
i \left[{\cal D}^{(0)}(q)\right]^{-1}_{\mu\nu} A^{A,\nu}_{q}
-\frac{1}{2}\left[A^{A,\mu}_{-q} -iq^{\mu}\phi^{A}_{-q} \right]
\Pi_{\mu\nu}(q)
\left[A^{A,\nu}_{q}+iq^{\nu}\phi^{A}_{q} \right] +\ldots,
\label{L-g-phi}
\ea
with the ellipsis denoting the interactions terms. Note that the presence
of the phase field octet $\phi^{A}_{q}$ in the last expression is very
important for preserving the gauge invariance of the model. Under a gauge
transformation, the phase $\phi^{A}_{q}$ in Eq.~(\ref{L-g-phi}) shifts
so that it exactly compensates the transformation of the gluon field.
Here we consider only infinitesimally small gauge transformations. In
general, the gauge transformation of $\phi^{A}_{q}$ field is not a simple
shift. However, it is always true that its transformation exactly
compensates the transformation of the gluon field.

The derivation of the effective Lagrangian density (\ref{L-g-phi}) in
dense QCD is not an easy task. In principle, one should start by
integrating out the high energy gluons which, as
is known from Refs.~\cite{PR1,Son,us,SW2,PR2,H1,Br1,us2,Spt,PR-weak}, play
the dominant role in the dynamical generation of the gap. As a result of
the integration, a new 4-fermion interaction term would be produced in the
action. This latter could be treated by introducing the quantum
Hubbard-Stratonovich composite field $\hat{\Delta}$ with the quantum
numbers of the color superconducting condensate. The effective Lagrangian
density in Eq.~(\ref{L-g-phi}), then, is given by integrating out the
quark degrees of freedom in the mean field approximation in which the
vacuum expectation value of the composite field is fixed constant. This
constant is identified with the value of the order parameter, $\langle
\hat{\Delta} \rangle \equiv \Delta$. The fields $\phi^{A}$ are the phase
fields of the composite field $\hat{\Delta}$ that correspond to the (would
be) NG bosons.

While the microscopic theory of dense quark matter may truly be
non-perturbative, the dynamics of gluon quasiparticles, given by
the Lagrangian density in Eq.~(\ref{L-g-phi}), could still be {\em
weakly} interacting. This is known to be the case at least in the
limit of asymptotically large quark densities. In this case, the
explicit expression for the polarization tensor $\Pi_{\mu\nu}$ can
be derived. Moreover, it is likely that the theory defined by
Eq.~(\ref{L-g-phi}) is applicable even at realistic densities
existing at the cores of compact stars (despite the fact that the
derivation of Eq.~(\ref{L-g-phi}) fails at low density). This is
because, in the case of weakly coupled gluons, the expression in
Eq.~(\ref{L-g-phi}) is the only possible quadratic expression
consistent with the $SU(3)_{c}$ color gauge symmetry.

In what follows, we use the simplest approximation for the polarization
tensor $\Pi_{\mu\nu}$ given by the improved hard dense loop (IHDL)
approximation. Here ``improved" indicates that the effect of the nonzero
values of quark gaps are taken into account in the calculation of loop
diagrams. The formal expression of such tensor was presented in
Ref.~\cite{Rsch2}. The zero temperature limit was also considered in
Ref.~\cite{Zar}. Below, we present some other limits which has not been
analyzed previously. In our notation, the explicit form of the
(left-handed contribution of) IHDL polarization tensor (in the
Matsubara formalism) reads
\ba
\Pi_{L,\mu\nu}(i\Omega_{m},q) &=& -\frac{g^{2}T}{24}\sum_{n}
\int\frac{d^{3}p}{(2\pi)^{3}} \mbox{tr}\left[
5 S_{8}(i\omega_{n},p)\gamma_{\mu}\gamma^{5}
S_{8}(i\omega_{n}+i\Omega_{m},p+q) \gamma_{\nu}\gamma^{5}
\right.\nonumber \\
&&\left.
+9 S_{8}(i\omega_{n},p) \gamma_{\mu}
S_{8}(i\omega_{n}+i\Omega_{m},p+q) \gamma_{\nu}
+2 S_{8}(i\omega_{n},p)\gamma_{\mu}\gamma^{5}
S_{1}(i\omega_{n}+i\Omega_{m},p+q)\gamma_{\nu}\gamma^{5}
\right.\nonumber \\
&&\left.
+2 S_{1}(i\omega_{n},p)\gamma_{\mu}\gamma^{5}
S_{8}(i\omega_{n}+i\Omega_{m},p+q)\gamma_{\nu}\gamma^{5}
\right],
\label{Pi-q}
\ea
where $\omega_{n}=(2n+1)\pi T$ and $\Omega_{m}=2m\pi T$ are the fermion
and boson Matsubara frequencies. In the derivation of Eq.~(\ref{Pi-q}), 
we used the identities (\ref{iden1}) and (\ref{iden2}) in
Appendix~\ref{AppA}. The right-handed contribution $\Pi_{R,\mu\nu}$ is
similar, and the full tensor is $\Pi_{\mu\nu} =\Pi_{L,\mu\nu}
+\Pi_{R,\mu\nu}$ (one could check, in fact, that
$\Pi_{R,\mu\nu}=\Pi_{L,\mu\nu}$).

Before going into further details, let us discuss the general properties
of the polarization tensor in the IHDL approximation given in
Eq.~(\ref{Pi-q}). We start by emphasizing that such a tensor, obtained by
integrating out the quark degrees of freedom, is {\em not} transverse in
general (neither is it directly related to the spectral density of gluons,
as we shall see later). This property is not connected with the
non-Abelian nature of the system at hand. Indeed, to the leading order,
dense QCD is essentially an Abelian theory, and many of its properties are
similar to those in QED \cite{bs-short-long}.
Having said
this, we should note right away that $\Pi_{\mu\nu}$ {\em is} transverse in
the normal phase of the quark matter (above the critical temperature
$T_{c}$). This is because the NG phase fields are absent in the normal
(symmetric) phase, and the polarization tensor is directly related to the
observable current-current correlation function.

In the broken phase (below $T_{c}$), on the other 
hand, a longitudinal part should necessarily appear. 
Of course, the existence of the longitudinal 
component in $\Pi_{\mu\nu}$ cannot spoil the gauge 
invariance of the model in Eq.~(\ref{L-g-phi}).
To clarify this point, we should emphasize 
that the one-loop quark contribution $\Pi_{\mu\nu}$ 
is only one of two contributions to the complete
expression for the gluon polarization tensor. As
we show in Sec.~\ref{c-c-function} [see, for example, 
Eq.~(\ref{L-g})], the other term comes from integrating
out the ``would be NG bosons" $\phi^{A}_{q}$. A simple
analysis shows that the final expression for the 
polarization tensor is in agreement with the 
Slavnov-Taylor identity required by gauge invariance. 
It is approapriate to mention that a similar observation
at the level of the effective theory was also made
in Ref.~\cite{C&D}.

As we show in Sec.~\ref{NG-boson} [see also the arguments 
of Ref.~\cite{Zar}], the longitudinal part of $\Pi_{\mu\nu}$
determines the dispersion relation of the pseudoscalar NG
bosons in the broken phase. 
Some explanation is in order here. Strictly speaking, the
properties of the NG bosons are related to the ``axial" polarization
tensor. In QCD at large density of quarks, this latter is approximately
equal to the polarization tensor in Eq.~(\ref{Pi-q}). Thus, by using the
leading order approximation, we could safely interchange the ``axial" and
``vector" polarization tensors.

In passing, we note that the IHDL approximation, 
utilized in this paper, is selfconsistent. As one could 
check, the IHDL gluon propagator considerably deviates
from the ordinary hard dense loop expression
only in the infrared region $q_{0},|\vec{q}| \alt
\Delta$. This is the region where the Meissner effect is 
seen and where low energy modes come into the game.
It is well known, however, that the relevant region of
momenta in the gap equation is $\Delta \ll q_{0}|,
\vec{q}| \ll \mu $. Thus, using either the IHDL or 
ordinary hard dense loop approximation would 
not affect the leading order solution to the gap
equation. At the same time, in order to preserve the
gauge invariance of the model (e.g., in order to
satisfy the Slavnov-Taylor identity) it is crucial
to use the IHDL approximation which, by construction, 
insures that the gauge invariance is intact.

\section{General structure of the one-loop polarization tensor}
\label{pol-tensor}

By substituting the explicit form of quark propagators, given in
Eqs.~(\ref{S-1}) and (\ref{S-2}), into the definition of the
polarization tensor in Eq.~(\ref{Pi-q}), we derive the general
result in Eq.~(\ref{Pi-most-general}) in Appendix~\ref{AppB}.
Then, by dropping all the terms suppressed by powers of the
chemical potential, we arrive at the following expression (here
both the left- and right-handed contributions are taken into
account):
\ba \Pi_{\mu\nu}(i\Omega_{m},q) &=&
\frac{g^{2}\mu^{2}}{2 \pi^{2}} \left(u_{\mu}
u_{\nu}-g_{\mu\nu}\right) +\frac{g^{2} \mu^{2}}{ 12 \pi^{2}}
T\sum_{n}\int_{-1}^{1} d \xi \int d \epsilon \nonumber \\
&\times&\Bigg\{
\left(u_{\mu} u_{\nu} (1-\xi^{2}) + \xi^{2} g_{\mu\nu}
+\frac{1-3\xi^{2}}{2} O^{(1)}_{\mu\nu}(q)\right)
\left[\frac{2 |\Delta_{T}|^{2}}
{(\omega_{n}^{2} + E_{-}^{2})
[(\omega_{n}+\Omega_{m})^{2} +E_{+}^{2}]}
\right. \nonumber \\
&&\left.
+\frac{2 |\Delta_{T}|^{2}}
{(\omega_{n}^{2} +E_{-}^{2})
[(\omega_{n}+\Omega_{m})^{2} +\tilde{E}_{+}^{2}]}
+\frac{2 |\Delta_{T}|^{2}}
{(\omega_{n}^{2} +\tilde{E}_{-}^{2})
[(\omega_{n}+\Omega_{m})^{2} +E_{+}^{2}]}
\right]
\nonumber \\
&+&\left(u_{\mu}u_{\nu}(1+\xi^{2})-\xi^{2}g_{\mu\nu}
-\frac{1-3\xi^{2}}{2} O^{(1)}_{\mu\nu}(q)\right)
\nonumber \\
&\times&
\left[
7\frac{\epsilon^{2}-\xi^{2} |\vec{q}|^{2}/4
-\omega_{n}(\omega_{n}+\Omega_{m})}
{(\omega_{n}^{2} +E_{-}^{2})
[(\omega_{n}+\Omega_{m})^{2} +E_{+}^{2}]}
+\frac{\epsilon^{2}-\xi^{2} |\vec{q}|^{2}/4
-\omega_{n}(\omega_{n}+\Omega_{m})}
{(\omega_{n}^{2} +E_{-}^{2})
[(\omega_{n}+\Omega_{m})^{2} +\tilde{E}_{+}^{2}]}
\right. \nonumber \\
&&\left.
+\frac{\epsilon^{2}-\xi^{2} |\vec{q}|^{2}/4
-\omega_{n}(\omega_{n}+\Omega_{m})}
{(\omega_{n}^{2} +\tilde{E}_{-}^{2})
[(\omega_{n}+\Omega_{m})^{2} +E_{+}^{2}]}
\right]
\nonumber \\
&+&\xi \frac{u_{\nu}\vec{q}_{\mu}+u_{\mu}\vec{q}_{\nu}}{|\vec{q}|}
\left[
7\frac{2 i\omega_{n} \epsilon
+ i\Omega_{m}(\epsilon-|\vec{q}|\xi/2)}
{(\omega_{n}^{2} + E_{-}^{2})
[(\omega_{n}+\Omega_{m})^{2} +E_{+}^{2}]}
\right. \nonumber \\
&&\left. +\frac{2 i\omega_{n} \epsilon +
i\Omega_{m}(\epsilon-|\vec{q}|\xi/2)} {(\omega_{n}^{2} +E_{-}^{2})
[(\omega_{n}+\Omega_{m})^{2} +\tilde{E}_{+}^{2}]} +\frac{2
i\omega_{n} \epsilon + i\Omega_{m}(\epsilon-|\vec{q}|\xi/2)}
{(\omega_{n}^{2} +\tilde{E}_{-}^{2}) [(\omega_{n}+\Omega_{m})^{2}
+E_{+}^{2}]} \right] \Bigg\} , \label{Pi-long}
\ea
where we use the notation $u_{\mu} = (1,0,0,0)$ and $\vec{q}_{\mu} =
q_{\mu} - (u\cdot q) u_{\mu}$.  Besides that, $O^{(1)}_{\mu\nu}(q)$ is one
of the operators defined in Appendix~\ref{AppZ}, see Eq.~(\ref{def-O1}).
This is the projector of the magnetic gluon modes. In Appendix~\ref{AppZ},
we also introduce the projectors of the electric and unphysical
(longitudinal in a 3+1 dimensional sense) modes, as well as intervening
operator $O^{(4)}_{\mu \nu}(q)$, mixing the electric and unphysical modes.  
It is appropriate to note here that the first term in Eq.~(\ref{Pi-long})
is the only term which got a contribution from the antiquark
quasiparticles. The other terms came from low energy quasiparticles
around the Fermi surface. These latter are characterized by the gap
$|\Delta_{T}| \equiv |\Delta^{-}_{T}|$ (see the discussion at the end of
Sec.~\ref{model}). In Eq.~(\ref{Pi-long}), we also use the
notation:
\ba
E_{\pm} &=&
\sqrt{(\epsilon\pm \xi |\vec{q}| /2)^{2}+|\Delta_{T}|^{2}} ,\\
\tilde{E}_{\pm} &=&
\sqrt{(\epsilon\pm \xi |\vec{q}| /2)^{2}+4|\Delta_{T}|^{2}},
\ea
which are the quasiparticle energies in the octet and singlet channels,
respectively. In the calculation, we subtracted all divergencies
independent of temperature and chemical potential. So, the result of
Eq.~(\ref{Pi-long}) is already given in terms of renormalized
quantities. Notice that the first term on the right hand side of
Eq.~(\ref{Pi-long}) comes from divergent antiquark-quark loop
contributions. In deriving the result in Eq.~(\ref{Pi-long}), we
performed the integration over the azimuthal angle, using the following
relations:
\ba
\frac{1}{2\pi}\int_{0}^{2\pi} d \varphi
\frac{\vec{p}^{\nu}}{|\vec{p}|} &=&
\xi \frac{\vec{q}^{\nu}}{|\vec{q}|}, \\
\frac{1}{2\pi}\int_{0}^{2\pi} d \varphi
\frac{\vec{p}^{\mu}\vec{p}^{\nu}}{|\vec{p}|^{2}} &=&
-\frac{1-\xi^{2}}{2}(g^{\mu\nu}-u^{\mu} u^{\nu})
-\frac{1-3\xi^{2}}{2}
\frac{\vec{q}^{\mu}\vec{q}^{\nu}}{|\vec{q}|^{2}}. 
\ea 
As was
already mentioned, in accordance with leading order approximation,
we dropped all the terms suppressed by powers of the chemical
potential in Eq.~(\ref{Pi-long}). For example, we used
\ba
\int\frac{d^{3}p}{(2\pi)^{3}}(\ldots) &\approx &
\frac{\mu^{2}}{(2\pi)^{2}}\int_{-1}^{1} d \xi
\int_{-\mu}^{\mu} d \epsilon (\ldots), \\
\epsilon^{-}_{p+q} \equiv |\vec{p}+\vec{q}|-\mu
&\approx & \epsilon^{-}_{p} + |\vec{q}| \xi + O(|\vec{q}|^2/\mu),
\ea
where $\xi$ is the cosine of the angle between $\vec{p}$ and $\vec{q}$.
Such an approximation is sufficient for studying the dynamics of the
collective modes with $|\Omega|, |\vec{q}| \ll \mu$. Similarly, we
approximated the results for the traces (with $|\vec{q}| \ll |\vec{p}|
\simeq \mu$)
\ba
\mbox{tr}\left[\gamma^{\mu} \Lambda^{(\pm)}_{p} {\cal P}_{\pm}
\gamma^{\nu} \Lambda^{(\pm)}_{p+q} {\cal P}_{\pm}\right]
&\simeq& g^{\mu\nu} - u^{\mu}u^{\nu}
+ \frac{\vec{p}^{\mu}\vec{p}^{\nu}}{|\vec{p}|^{2}},
\label{tr1++}\\
\mbox{tr}\left[\gamma^{\mu} \Lambda^{(\pm)}_{p} {\cal P}_{\pm}
\gamma^{\nu} \Lambda^{(\mp)}_{p+q} {\cal P}_{\mp}\right]
&\simeq& u^{\mu}u^{\nu}
-\frac{\vec{p}^{\mu}\vec{p}^{\nu}}{|\vec{p}|^{2}},
\label{tr1+-}\\
\mbox{tr}\left[\gamma^{\mu} \gamma^0 \Lambda^{(\pm)}_{p}
{\cal P}_{\pm}
\gamma^{\nu} \gamma^0 \Lambda^{(\pm)}_{p+q}{\cal P}_{\pm}\right]
&\simeq& \left(u^{\mu} \mp \frac{\vec{p}^{\mu}}{|\vec{p}|}\right)
\left(u^{\nu} \mp \frac{\vec{p}^{\nu}}{|\vec{p}|}\right),
\label{tr2++} \\
\mbox{tr}\left[\gamma^{\mu} \gamma^0 \Lambda^{(\pm)}_{p} {\cal
P}_{\pm} \gamma^{\nu} \gamma^0 \Lambda^{(\mp)}_{p+q} {\cal
P}_{\mp}\right] &\simeq& - g^{\mu\nu} + u^{\mu} u^{\nu}
-\frac{\vec{p}^{\mu}\vec{p}^{\nu}}{|\vec{p}|^{2}} .
\label{tr2+-}
\ea
The sum over the Matsubara frequencies in Eq.~(\ref{Pi-long})
is straightforward. It reduces to three types of simple sums given
in Eqs.~(\ref{def-F1}) -- (\ref{def-F3}) in Appendix~\ref{AppA}.
By making use of the summation formulas also given in
Appendix~\ref{AppA}, we derive another representation for the
polarization tensor given in Eq.~(\ref{Pi-after-T}). As is easy to
check, then, the Lorentz structure of $\Pi_{\mu\nu}(q)$ allows a
well defined decomposition of the result in terms of the $O^{(i)}$
operators, defined in Appendix~\ref{AppZ},
\be \Pi_{\mu\nu}(q) =
 \Pi_{1}(q)  O^{(1)}_{\mu\nu}(q)
+\Pi_{2}(q)  O^{(2)}_{\mu\nu}(q)
+\Pi_{3}(q)  O^{(3)}_{\mu\nu}(q)
+\Pi_{4}(q)  O^{(4)}_{\mu\nu}(q) .
\label{Pi-gen}
\ee
After a tedious, but straightforward rearrangement of different terms
in the polarization tensor, we arrive at the following explicit results
for the component functions:
\begin{mathletters}
\ba
\Pi_{1}(q) &=& \omega_{p}^{2} H(q)
\label{P1-HKLM}, \\
\Pi_{2}(q) &=& \frac{\omega_{p}^{2}}
{q_{0}^{2} -|\vec{q}|^{2}} \left[
- |\vec{q}|^{2} K(q)
+ q_{0}^{2} L(q)
+ 2 q_{0} |\vec{q}| M(q) \right]
\label{P2-HKLM}, \\
\Pi_{3}(q) &=& \frac{\omega_{p}^{2}}
{q_{0}^{2} -|\vec{q}|^{2}} \left[
q_{0}^{2} K(q)
- |\vec{q}|^{2} L(q)
- 2 q_{0} |\vec{q}| M(q) \right]
\label{P3-HKLM}, \\
\Pi_{4}(q) &=& \frac{\omega_{p}^{2}}
{q_{0}^{2} -|\vec{q}|^{2}} \left[
q_{0} |\vec{q}| K(q)
-q_{0} |\vec{q}|  L(q)
-(q_{0}^{2}+|\vec{q}|^{2}) M(q) \right] ,
\label{P4-HKLM}
\ea
\label{Pi-HKLM}
\end{mathletters}
where $\omega_{p}\equiv g\mu/\sqrt{2}\pi$ is the plasma frequency. The
dimensionless scalar functions $H(q)$, $K(q)$, $L(q)$ and $M(q)$ on the
right hand side have the following representation:
\ba 
H(q) &=& -1 +\frac{1}{12}\int_{-1}^{1} d \xi
(1-\xi^{2}) \int d \epsilon  \nonumber \\
&\times & \Bigg[
\left(2|\Delta_{T}|^{2} +7 E_{-} E_{+}
-7\epsilon^{2} + 7\frac{|\vec{q}|^{2}\xi^{2}}{4} \right)
X(E_{-},E_{+})
+\left(2|\Delta_{T}|^{2} -7 E_{-} E_{+}
-7\epsilon^{2} + 7\frac{|\vec{q}|^{2}\xi^{2}}{4} \right)
Y(E_{-},E_{+})
\nonumber \\
&&+\left(2|\Delta_{T}|^{2} + E_{-} \tilde{E}_{+}
-\epsilon^{2} + \frac{|\vec{q}|^{2}\xi^{2}}{4} \right)
X(E_{-},\tilde{E}_{+})
+\left(2|\Delta_{T}|^{2} -E_{-} \tilde{E}_{+}
-\epsilon^{2} + \frac{|\vec{q}|^{2}\xi^{2}}{4} \right)
Y(E_{-},\tilde{E}_{+})
\nonumber \\
&&+\left(2|\Delta_{T}|^{2} + \tilde{E}_{-} E_{+}
-\epsilon^{2} + \frac{|\vec{q}|^{2}\xi^{2}}{4} \right)
X(\tilde{E}_{-},E_{+})
+\left(2|\Delta_{T}|^{2} -\tilde{E}_{-} E_{+}
-\epsilon^{2} + \frac{|\vec{q}|^{2}\xi^{2}}{4} \right)
Y(\tilde{E}_{-},E_{+})
\label{H-long} ,\\
K(q) &=& \frac{1}{6}\int_{-1}^{1} d \xi
\int d \epsilon    \nonumber \\
&\times & \Bigg[
\left(2|\Delta_{T}|^{2} -7 E_{-} E_{+}
+7\epsilon^{2} - 7\frac{|\vec{q}|^{2}\xi^{2}}{4} \right)
X(E_{-},E_{+})
+\left(2|\Delta_{T}|^{2} +7 E_{-} E_{+}
+7\epsilon^{2} - 7\frac{|\vec{q}|^{2}\xi^{2}}{4} \right)
Y(E_{-},E_{+})
\nonumber \\
&&+\left(2|\Delta_{T}|^{2} - E_{-} \tilde{E}_{+}
+\epsilon^{2} - \frac{|\vec{q}|^{2}\xi^{2}}{4} \right)
X(E_{-},\tilde{E}_{+})
+\left(2|\Delta_{T}|^{2} +E_{-} \tilde{E}_{+}
+\epsilon^{2} - \frac{|\vec{q}|^{2}\xi^{2}}{4} \right)
Y(E_{-},\tilde{E}_{+})
\nonumber \\
&&+\left(2|\Delta_{T}|^{2} - \tilde{E}_{-} E_{+}
+\epsilon^{2} - \frac{|\vec{q}|^{2}\xi^{2}}{4} \right)
X(\tilde{E}_{-},E_{+})
+\left(2|\Delta_{T}|^{2} +\tilde{E}_{-} E_{+}
+\epsilon^{2} - \frac{|\vec{q}|^{2}\xi^{2}}{4} \right)
Y(\tilde{E}_{-},E_{+})
\label{K-long} ,\\
L(q) &=& - 1 + \frac{1}{6}\int_{-1}^{1} \xi^{2}
d \xi \int d \epsilon  \nonumber \\
&\times & \Bigg[
\left(2|\Delta_{T}|^{2} +7 E_{-} E_{+}
-7\epsilon^{2} + 7\frac{|\vec{q}|^{2}\xi^{2}}{4} \right)
X(E_{-},E_{+})
+\left(2|\Delta_{T}|^{2} -7 E_{-} E_{+}
-7\epsilon^{2} + 7\frac{|\vec{q}|^{2}\xi^{2}}{4} \right)
Y(E_{-},E_{+})
\nonumber \\
&&+\left(2|\Delta_{T}|^{2} + E_{-} \tilde{E}_{+}
-\epsilon^{2} + \frac{|\vec{q}|^{2}\xi^{2}}{4} \right)
X(E_{-},\tilde{E}_{+})
+\left(2|\Delta_{T}|^{2} -E_{-} \tilde{E}_{+}
-\epsilon^{2} + \frac{|\vec{q}|^{2}\xi^{2}}{4} \right)
Y(E_{-},\tilde{E}_{+})
\nonumber \\
&&+\left(2|\Delta_{T}|^{2} + \tilde{E}_{-} E_{+}
-\epsilon^{2} + \frac{|\vec{q}|^{2}\xi^{2}}{4} \right)
X(\tilde{E}_{-},E_{+})
+\left(2|\Delta_{T}|^{2} -\tilde{E}_{-} E_{+}
-\epsilon^{2} + \frac{|\vec{q}|^{2}\xi^{2}}{4} \right)
Y(\tilde{E}_{-},E_{+})
\label{L-long},\\
M(q) &=& \frac{1}{6}\int_{-1}^{1} d \xi
\int d \epsilon  \nonumber \\
&\times & \Bigg[
7 q_{0} \left( \epsilon \xi \frac{E_{+}-E_{-}}{E_{+}+E_{-}}
-\frac{|\vec{q}|\xi^{2}}{2} \right) X(E_{-},E_{+})
+7 q_{0} \left( \epsilon \xi \frac{E_{+}+E_{-}}{E_{+}-E_{-}}
-\frac{|\vec{q}|\xi^{2}}{2} \right) Y(E_{-},E_{+})
\nonumber \\
&& +q_{0} \left( \epsilon \xi \frac{\tilde{E}_{+}-E_{-}}
{\tilde{E}_{+}+E_{-}}
-\frac{|\vec{q}|\xi^{2}}{2} \right) X(E_{-},\tilde{E}_{+})
+ q_{0} \left( \epsilon \xi \frac{\tilde{E}_{+}+E_{-}}
{\tilde{E}_{+}-E_{-}}
-\frac{|\vec{q}|\xi^{2}}{2} \right) Y(E_{-},\tilde{E}_{+})
\nonumber \\
&&  +q_{0} \left( \epsilon \xi \frac{E_{+}-\tilde{E}_{-}}
{E_{+}+\tilde{E}_{-}}
-\frac{|\vec{q}|\xi^{2}}{2} \right) X(\tilde{E}_{-},E_{+})
+ q_{0} \left( \epsilon \xi \frac{E_{+}+\tilde{E}_{-}}
{E_{+}-\tilde{E}_{-}}
-\frac{|\vec{q}|\xi^{2}}{2} \right) Y(\tilde{E}_{-},E_{+})
\Bigg]  ,
\label{M-long}
\ea
where we introduced the following two functions:
\ba
X(a,b) = \frac{(a+b)[1-n(a)-n(b)]}{2ab [(a+b)^{2}-q_{0}^{2}]},
&\qquad&
Y(a,b) = \frac{(a-b)[n(a)-n(b)]}{2ab [(a-b)^{2}-q_{0}^{2}]},
\ea
and $n(a)$ and $n(b)$ are the Fermi distribution functions.

\section{Current-current correlation function and propagators}
\label{c-c-function}

The representation of the polarization tensor in
Eq.~(\ref{Pi-gen}) in terms of three projectors and one
intervening operator allows us to give a rather general treatment
of the collective modes coupled to the vector and axial color
currents.  Before studying the properties of the collective modes
in detail, it is instructive to consider the structure of the
current-current correlation function in terms of the component
functions of the polarization tensor $\Pi_{i}(q)$, see the
definition in Eq.~(\ref{Pi-gen}).

Let us start with the Lagrangian density for the gauge field in
Eq.~(\ref{L-g-phi}). In order to be able to generate the current-current
correlation function, we introduce the classical source terms
$A_{cl}^{A,\mu}(q)$ into the action,
\ba
{\cal L}_{g,\phi} &=& -\frac{1}{2} A^{A,\mu}(-q) q^{2} \left[
O^{(1)}_{\mu\nu}(q) +O^{(2)}_{\mu\nu}(q)
+\lambda O^{(3)}_{\mu\nu}(q) \right] A^{A,\nu}(q)
\nonumber \\
&&-\frac{1}{2}\left[
A_{cl}^{A,\mu}(-q) + A^{A,\mu}(-q) -i q^{\mu}\phi^{A}(-q)
\right] \Pi_{\mu\nu}(q) \left[
A_{cl}^{A,\nu}(q)+ A^{A,\nu}(q) + i q^{\nu}\phi^{A}(q)
\right] +\ldots ,
\label{L-g-pi}
\ea
where the ellipsis denote the interactions terms.  Notice that the
classical sources enter only through the second term of the Lagrangian
density which is generated by integrating out the quarks. The gauge fixing
parameter is denoted by $\lambda$. As usual, it comes along with the
projector of the unphysical gluon modes. The phase fields $\phi^{A}$,
required by gauge symmetry, enter the action in the same way as the
longitudinal components of the gauge fields. Now, in order to derive the
action for the transverse (physical) gluons which contain the information
about the current-current correlation functions, we should integrate out
the phase fields $\phi^{A}$. Thus, we obtain the following expression:
\ba
{\cal L}_{g} &=& -\frac{1}{2} A^{A,\mu}(-q) q^{2} \left[
O^{(1)}_{\mu\nu}(q) +O^{(2)}_{\mu\nu}(q)
+\lambda O^{(3)}_{\mu\nu}(q)
\right] A^{A,\nu}(q) \nonumber \\
&&- \frac{1}{2}\left[ A_{cl}^{A,\mu}(-q) + A^{A,\mu}(-q) \right]
\left[ \Pi_{\mu\nu}(q)
- \frac{\Pi_{\mu\lambda} q^{\lambda}
q^{\kappa} \Pi_{\kappa\nu}}
{q_{\rho} \Pi^{\rho\sigma} q_{\sigma}}
\right] \left[
A_{cl}^{A,\nu}(q) + A^{A,\nu}(q) \right]
+\ldots .
\label{L-g}
\ea
By substituting the decomposition of the polarization tensor
$\Pi_{\mu\nu}$ in Eq.~(\ref{Pi-gen}), we arrive at the action:
\ba
{\cal L}_{g} &=& -\frac{1}{2}A^{A,\mu}(-q) q^{2} \left[
O^{(1)}_{\mu\nu}(q) +O^{(2)}_{\mu\nu}(q)
+\lambda O^{(3)}_{\mu\nu}(q)
\right] A^{A,\nu}(q) \nonumber \\
&&- \frac{1}{2}\left[ A_{cl}^{A,\mu}(-q) + A^{A,\mu}(-q) \right]
\left[ \Pi_{1}(q) O^{(1)}_{\mu\nu}(q) + \left(\Pi_{2}(q)
+\frac{[\Pi_{4}(q)]^{2}}{\Pi_{3}(q)} \right)O^{(2)}_{\mu\nu}(q)
\right] \left[ A_{cl}^{A,\nu}(q) + A^{A,\nu}(q) \right] +\ldots .
\label{L-g-1}
\ea
Finally, by integrating out the quantum gauge fields, we arrive at the
following quadratic form:
\ba
{\cal L}_{cl} &=&  \frac{1}{2}A_{cl}^{A,\mu}(-q) \left[ \frac{q^{2}
\Pi_{1}}{q^{2}+\Pi_{1}} O^{(1)}_{\mu\nu}(q) + \frac{q^{2}
\left[\Pi_{2} \Pi_{3} + (\Pi_{4})^{2}\right]} {(q^{2}+ \Pi_{2})
\Pi_{3} + (\Pi_{4})^{2} } O^{(2)}_{\mu\nu}(q) \right]
A_{cl}^{A,\nu}(q).
\label{L-cl}
\ea
As is clear, the tensor in the square brackets of this quadratic form
defines the current-current correlation function, i.e.,
\be
\langle j^{A}_{\mu} j^{B}_{\nu}
\rangle_{q} = \delta^{AB} \left[ \frac{q^{2}
\Pi_{1}}{q^{2}+\Pi_{1}} O^{(1)}_{\mu\nu}(q) + \frac{q^{2}
\left[\Pi_{2} \Pi_{3} + (\Pi_{4})^{2}\right]} {(q^{2}+ \Pi_{2})
\Pi_{3} + (\Pi_{4})^{2} } O^{(2)}_{\mu\nu}(q) \right].
\label{cor-function} 
\ee 
The location of its poles, 
\ba 
q^{2} +
\Pi_{1}(q) &=& 0, \quad \mbox{``magnetic"},
\label{spectrum-mag}\\
\left[q^{2} + \Pi_{2}(q)\right] \Pi_{3}(q) + [\Pi_{4}(q)]^{2} &=&
0, \quad \mbox{``electric"},
\label{spectrum-el}
\ea
defines (in a gauge invariant way) the spectrum of collective
excitations in the model at hand. It is interesting to point out
that one could have extracted this spectrum of the collective
modes already from the poles of the quantum gauge field propagator
in Eq.~(\ref{L-g-1}): \be {\cal D}^{(g)}_{\mu\nu} (q)=
-\frac{i}{q^{2}+\Pi_{1}} O^{(1)}_{\mu\nu}(q)
-\frac{i\Pi_{3}}{(q^{2}+\Pi_{2})\Pi_{3}+(\Pi_{4})^{2}}
  O^{(2)}_{\mu\nu}(q)
-\frac{i}{\lambda q^{2}} O^{(3)}_{\mu\nu}(q).
\ee
We performed the last integration just to show explicitly that the proper
definition of the current-current correlation function indeed gives the
anticipated result.

Before concluding this section, it is also instructive to perform the
following exercise. Let us perform the integration in Eq.~(\ref{L-g-pi})
over the quantum gauge field first.  Then, the intermediate action for
the ``phase" field reads:
\ba
{\cal L}_{\phi} &=&
-\frac{1}{2}\left[ A_{cl}^{A,\mu}(-q) -iq^{\mu}\pi^{A}(-q) \right]
\left[
\frac{q^{2} \Pi_{1}}{q^{2} + \Pi_{1}} O^{(1)}_{\mu\nu}(q)
+\frac{q^{2}
\left[(\lambda q^{2} + \Pi_{3})\Pi_{2}+(\Pi_{4})^{2}\right]}
{\lambda q^{2} (q^{2} + \Pi_{2})
+\left[(q^{2} + \Pi_{2})\Pi_{3}+(\Pi_{4})^{2}\right]}
O^{(2)}_{\mu\nu}(q)
\right. \nonumber \\
&& \left.
+\frac{\lambda q^{2}
\left[(q^{2} + \Pi_{2})\Pi_{3}+(\Pi_{4})^{2}\right]}
{\lambda q^{2} (q^{2} + \Pi_{2})
+\left[(q^{2} + \Pi_{2})\Pi_{3}+(\Pi_{4})^{2}\right]}
O^{(3)}_{\mu\nu}(q)
\right. \nonumber \\
&& \left.
+\frac{\lambda (q^{2})^{2} \Pi_{4}}
{\lambda q^{2} (q^{2} + \Pi_{2})
+\left[(q^{2} + \Pi_{2})\Pi_{3}+(\Pi_{4})^{2}\right]}
O^{(4)}_{\mu\nu}(q)
\right]
\left[ A_{cl}^{A,\nu}(q)+ iq^{\nu}\pi^{A}(q) \right]
+\ldots .
\label{L-phi}
\ea
At first glance, it does not make sense to put any physical relevance
behind the propagator of the $\phi^{A}$ field, because the latter is a
gauge dependent phase field. In condensed matter physics, however,
this type of propagator is often used to extract the spectrum of the
electric-type collective modes, such as the CG modes. The explicit 
form of the $\phi^{A}$ field propagator reads
\be
{\cal D}^{(\phi)}(q) =
-i\frac{\lambda q^{2}(q^{2}+\Pi_{2})
+[(q^{2}+\Pi_{2})\Pi_{3}+(\Pi_{4})^{2}]}
{\lambda (q^{2})^{2}[(q^{2}+\Pi_{2})\Pi_{3}+(\Pi_{4})^{2}]}.
\label{propagator-phi}
\ee
We see that the location of one of the poles gives exactly the same
spectrum as that in Eq.~(\ref{spectrum-el}). It is remarkable that the
position of the corresponding pole and its residue are gauge invariant.
However, there exists another pole in the propagator in
Eq.~(\ref{propagator-phi}) at $q^{2}=0$. This latter is unphysical in a
sense that it does not appear in the current-current correlation function,
given in Eq.~(\ref{L-cl}). The appearance of the unphysical pole clearly
shows that the use of the propagator (\ref{propagator-phi}) for extracting
spectra of collecting modes is limited. For completeness, let us also
mention that there is no trace of the magnetic collective modes in the
propagator of the longitudinal $\phi^{A}$ field. The information about the
magnetic modes, however, is not lost in Eq.~(\ref{L-phi}). As is easy to
check, after integrating out the $\phi^{A}$ field in Eq.~(\ref{L-phi}), we
arrive at the same result as in Eq.~(\ref{L-cl}).

\section{Collective modes: general formalism}
\label{Spectra}

In the preceding section, we derived the general expression for the
current-current correlation function, see Eq.~(\ref{cor-function}). The
location of its poles gives the spectra of the electric and magnetic type
collective modes (assuming that their width is not too large). The
explicit form of such spectra is determined by Eqs.~(\ref{spectrum-mag})
and (\ref{spectrum-el}). By making use of the definition in
Eq.~(\ref{Pi-HKLM}), the electric and magnetic dispersion relations
take the following form:
\ba
q_{0}^{2}-|\vec{q}|^{2} + \omega_{p}^{2}H(q) &=& 0,
\quad \mbox{``magnetic"}
\label{spec-mag}, \\
q_{0}^{2} K(q)
-|\vec{q}|^{2} L(q)
-2 q_{0} |\vec{q}| M(q)
+\omega_{p}^{2}\left[  K(q)L(q) +M^{2}(q) \right] &=& 0,
\quad \mbox{``electric"}.
\label{spec-el}
\ea
Here we used the following identity, relating different
component functions:
\be
\Pi_{2} \Pi_{3} +(\Pi_{4})^{2} = \omega_{p}^{4}
\left[ K L + M^{2} \right].
\ee
To the leading order in powers of $q_{0}$ and $|\vec{q}|$ (keeping the
ratio $v=q_{0}/|\vec{q}|$ arbitrary and $|\Delta_{T}|$ finite), we
obtain the following result for the component functions of the polarization
tensor:
\ba
H(q) &=& -1 + \frac{1}{9} \int d \epsilon  \Bigg\{
\frac{9|\Delta_{T}|^{2}\tanh(\beta E/2)}{4 E^{3}}
- \frac{(5|\Delta_{T}|^{2}+14 \epsilon^{2})n^{\prime}(E)}{2E^{2}}
\left[1-\frac{3 E^{2} q_{0}^{2}}{2\epsilon^{2}|\vec{q}|^{2}}
-\frac{3Eq_{0}}{4\epsilon|\vec{q}|}\left(1
-\frac{ E^{2} q_{0}^{2}}{\epsilon^{2}|\vec{q}|^{2}}
\right)\ln \frac{E q_{0}+\epsilon|\vec{q}|}
{Eq_{0}-\epsilon|\vec{q}|}\right] \nonumber \\
&&+\frac{(2|\Delta_{T}|^{2}-\epsilon^{2}+\tilde{E}E)
(E+\tilde{E})[1-n(\tilde{E})-n(E)]}
{\tilde{E} E (E+\tilde{E})^{2}}
+\frac{(2|\Delta_{T}|^{2}-\epsilon^{2}-\tilde{E}E)
(\tilde{E}-E)[n(\tilde{E})-n(E)]}
{E \tilde{E} (\tilde{E}-E)^{2}}
\Bigg\}
\label{H-short} ,\\
K(q) &=& \frac{1}{3} \int d \epsilon  \Bigg[
-\frac{5|\Delta_{T}|^{2}\tanh(\beta E/2)}{4 E^{3}}
+\frac{(9|\Delta_{T}|^{2}+14 \epsilon^{2})n^{\prime}(E)}{2E^{2}}
\left(1-\frac{E q_{0}}{2\epsilon|\vec{q}|}
\ln \frac{E q_{0}+\epsilon|\vec{q}|}
{Eq_{0}-\epsilon|\vec{q}|}\right) \nonumber \\
&&+\frac{(2|\Delta_{T}|^{2}+\epsilon^{2}-\tilde{E}E)
(E+\tilde{E})[1-n(\tilde{E})-n(E)]}
{\tilde{E} E (E+\tilde{E})^{2}}
+\frac{(2|\Delta_{T}|^{2}+\epsilon^{2}+\tilde{E}E)
(\tilde{E}-E)[n(\tilde{E})-n(E)]}
{E \tilde{E} (\tilde{E}-E)^{2}}
\Bigg],
\label{K-short}\\
L(q) &=& -1 + \frac{1}{9} \int d \epsilon  \Bigg[
\frac{9|\Delta_{T}|^{2}\tanh(\beta E/2)}{4 E^{3}}
- \frac{(5|\Delta_{T}|^{2}+14 \epsilon^{2})n^{\prime}(E)}{2E^{2}}
\left(1+\frac{3 E^{2} q_{0}^{2}}{\epsilon^{2}|\vec{q}|^{2}}
-\frac{3E^{3} q_{0}^{3}}{2\epsilon^{3}|\vec{q}|^{3}}\ln
\frac{E q_{0}+\epsilon|\vec{q}|}
{Eq_{0}-\epsilon|\vec{q}|}\right) \nonumber \\
&&+\frac{(2|\Delta_{T}|^{2}-\epsilon^{2}+\tilde{E}E)
(E+\tilde{E})[1-n(\tilde{E})-n(E)]}
{\tilde{E} E (E+\tilde{E})^{2}}
+\frac{(2|\Delta_{T}|^{2}-\epsilon^{2}-\tilde{E}E)
(\tilde{E}-E)[n(\tilde{E})-n(E)]}
{E \tilde{E} (\tilde{E}-E)^{2} }
\Bigg],
\label{L-short}\\
M(q) &=& \frac{7}{3}\frac{q_{0}}{|\vec{q}|}
\int d \epsilon  \left(
1-\frac{E q_{0}}{2\epsilon |\vec{q}|}
\ln \frac{E q_{0}+\epsilon |\vec{q}|}
{E q_{0}-\epsilon |\vec{q}|}\right) n^{\prime}(E),
\label{M-short}
\ea
where the quasiparticle energies in the octet and singlet channels
are $E=\sqrt{\epsilon^{2}+|\Delta_{T}|^{2}}$ and
$\tilde{E}=\sqrt{\epsilon^{2}+4|\Delta_{T}|^{2}}$, respectively,
and $\beta\equiv 1/T$ is the inverse temperature.

For further convenience, we rewrite these representations in the
following shorthand form:
\begin{mathletters}
\ba
H(v) &=& -1 +\frac{1}{9}\left[9I_{1}+I_{3}+I_{5}-I_{9}(v)\right] ,
\label{H-Is}\\
K(v) &=& \frac{1}{3}\left[-5I_{1}+I_{2}+I_{4}+I_{6}(v)\right] ,
\label{K-Is} \\
L(v) &=& -1+\frac{1}{9}\left[9I_{1}+I_{3}+I_{5}-I_{7}(v)\right] ,
\label{L-Is} \\
M(v) &=& \frac{7v}{3}I_{8}(v) , \label{M-Is}
\ea
\label{HKLM-Is}
\end{mathletters}
where the $I_{i}$ functions on the right hand side are defined in
Appendix~\ref{AppC}. In the same Appendix, we also present their
nearcritical asymptotes which include all corrections up to the second
power in $|\Delta_{T}|/T$.

\section{Plasmons}
\label{plasmon}

Dense quark matter is a non-Abelian plasma. Similarly to the ordinary
plasma, it is characterized by the so-called plasma frequency
$\omega_{p}$. By definition, the properties of the gluon field excitations
change qualitatively when the frequency increases from the values less
than $\omega_{p}$ to the values greater than $\omega_{p}$. For example, in
ordinary plasma (in absence of magnetic fields), only the modes with
frequencies greater than the plasma frequency could freely propagate in
the bulk. Such modes are called plasmons.

To derive the spectrum of the plasmon excitations, it is sufficient to
consider only the long wave length limit $|\vec{q}|\to 0$, keeping $q_{0}$
large. The only restriction is that $q_{0} \ll \mu$. As we shall see,
such an assumption is justified in the weakly coupled limit.

Let us start our analysis by writing down the component functions,
defined in Eqs.~(\ref{H-long}) -- (\ref{M-long}), in the long wave
length limit. By substituting $|\vec{q}| = 0$, we obtain
\ba
H(q_{0}) &=& L(q_{0}) = -1 + \frac{1}{9} \int d \epsilon  \Bigg[
\frac{9|\Delta_{T}|^{2} \tanh(\beta E/2) }{E(4 E^{2}-q_{0}^{2})}
+\frac{(2|\Delta_{T}|^{2}-\epsilon^{2}+\tilde{E}E)
(E+\tilde{E})[1-n(\tilde{E})-n(E)]}
{\tilde{E}E[(E+\tilde{E})^{2}-q_{0}^{2}]} \nonumber \\
&+& \frac{(2|\Delta_{T}|^{2}-\epsilon^{2}-\tilde{E}E)
(\tilde{E}-E)[n(\tilde{E})-n(E)]}{\tilde{E}E[(\tilde{E}-E)^{2}-q_{0}^{2}]}
\Bigg],
\label{H-T} \\
K(q_{0}) &=& \frac{1}{3} \int d \epsilon  \Bigg[
-\frac{5|\Delta_{T}|^{2} \tanh(\beta E/2)}{E(4 E^{2}-q_{0}^{2})}
+\frac{(2|\Delta_{T}|^{2}+\epsilon^{2}-\tilde{E}E)
(E+\tilde{E})[1-n(\tilde{E})-n(E)]}
{\tilde{E}E[(E+\tilde{E})^{2}-q_{0}^{2}]}\nonumber \\
&+& \frac{(2|\Delta_{T}|^{2}+\epsilon^{2}+\tilde{E}E)
(\tilde{E}-E)[n(\tilde{E})-n(E)]}
{\tilde{E}E[(\tilde{E}-E)^{2}-q_{0}^{2}]}
\Bigg],
\label{K-T} \\
M(q_{0}) &=& 0. 
\label{M-T} 
\ea 
It is straightforward to check that $H(q_{0})=L(q_{0})\simeq -1$ and
$K(q_{0})\simeq 0$ at large $q_{0}\gg |\Delta_{T}|$ (here we dropped all
real and imaginary contributions, suppressed by an inverse power of
$q_{0}^{2}$). In fact, these asymptotes are independent of the specific
values of the temperature and the color superconducting gap, provided that
the inequality $T,|\Delta_{T}| \ll q_{0}$ is satisfied. In this limit,
therefore, the dispersion relations for both the magnetic and electric
collective modes, see Eq.~(\ref{spec-mag}) and (\ref{spec-el}), take the
following form:
\be 
q_{0}^{2} = \omega_{p}^{2}, \quad \mbox{for} \quad
T,|\Delta_{T}| \ll q_{0} \ll \mu. 
\label{Omega=omega_p} 
\ee
where $\omega_{p}\equiv g \mu /\sqrt{2}\pi$ plays the role of the
plasma frequency. As we stated earlier, this result is in
agreement with the assumptions used.

For completeness of presentation, let us also write down the expressions
for the component functions of the polarization tensor in limit of large
$q_{0}$ (and $|\vec{q}|=0$). By making use of the definitions in
Eq.~(\ref{Pi-HKLM}), we derive
\ba
\Pi_{1}(q_{0}) &=& \Pi_{2}(q_{0}) \simeq \omega_{p}^{2} , \\
\Pi_{3}(q_{0}) &=& \Pi_{4}(q_{0}) \simeq 0.
\ea
The existence of the plasmon collective modes with large frequency is
hardly a surprise in the dense quark matter. As in the case of ordinary
metals, the plasmons are present in the normal and the color
superconducting phases, and their properties are weakly affected by a
small (compared to the chemical potential) temperature of the system and
by the presence of a gap. The ``light" plasmons, that we reveal and study
in this section too (see Subsec.~\ref{light-plasmon}), are much more
interesting collective modes of the color-flavor locked phase of dense
quark matter. In contrast to the ordinary plasmons, there is no trace of
such ``light" plasmon modes in the normal phase. Before studying them in
detail, it makes sense, therefore, to review the properties of all
collective modes in the normal phase first.

\subsection{Normal phase, $T \ll \mu$}
\label{Tllmu}

In the normal phase, the properties of the collective modes are well
known. Here we just briefly review them.

By definition, the value of the gap in the quasiparticle spectrum around
the Fermi surface is zero in the normal phase. Thus, we substitute
$|\Delta_{T}| = 0$ into our general expressions for the component
functions in Eqs.~(\ref{H-long}) -- (\ref{M-long}). By further
approximating the expressions in the limit of low temperatures
($T \ll \mu$) and small momenta ($|q_{0}|,|\vec{q}| \ll \mu$), we
arrive at the following results:
\ba
H(q) &=& - \frac{3}{2} \left[1 +
\frac{q_{0}^{2}-|\vec{q}|^{2}}{|\vec{q}|^{2}} \left(
1-\frac{q_{0}}{2|\vec{q}|}\ln \frac{q_{0}+|\vec{q}|}
{q_{0}-|\vec{q}|}\right) \right] , \\
K(q) &=&  -3 \left(
1-\frac{q_{0}}{2|\vec{q}|}\ln \frac{q_{0}+|\vec{q}|}
{q_{0}-|\vec{q}|}\right), \\
L(q) &=& \frac{3q_{0}^{2}}{|\vec{q}|^{2}}
\left(1-\frac{q_{0}}{2|\vec{q}|}\ln \frac{q_{0}+|\vec{q}|}
{q_{0}-|\vec{q}|}\right) , \\
M(q) &=& -\frac{3q_{0}}{|\vec{q}|} \left(
1-\frac{q_{0}}{2|\vec{q}|}\ln \frac{q_{0}+|\vec{q}|}
{q_{0}-|\vec{q}|}\right)  .
\ea
In the derivation, the ratio $q_{0}/|\vec{q}|$ was kept arbitrary.
Now, by making use of the definitions in Eq.~(\ref{Pi-HKLM}), we derive
the well known hard-dense loop expression for the polarization tensor
\cite{Heinz,Vija,Manuel}:
\ba
\Pi_{1}(q) &=& -\Pi_{t}(q) , \\
\Pi_{2}(q) &=& -\frac{q_{0}^{2}-|\vec{q}|^2}
{|\vec{q}|^2} \Pi_{l}(q), \\
\Pi_{3}(q) &=& \Pi_{4}(q) =0,
\ea
given in the same representation as in Ref.~\cite{us}, with
\ba
\Pi_{l}(q) &=& 3\omega_{p}^2
\left(\frac{q_{0}}{2|\vec{q}|}
\ln\left|\frac{q_{0}+|\vec{q}|}{q_{0}-|\vec{q}|}\right|-1
-i\pi\frac{q_{0}}{2|\vec{q}|}\theta(|\vec{q}|^2-q_{0}^{2})
\right), \\
\Pi_{t}(q) &=& \frac{3}{2}
\omega_{p}^2-\frac{q_{0}^{2}-|\vec{q}|^2}
{2|\vec{q}|^2} \Pi_{l}(q).
\ea
In accordance with the approximation ($T \ll \mu$), the finite temperature
corrections were neglected here. Such corrections are suppressed by at
least a factor of $(T/\mu)^{2}$. When the temperature is not small, one
would need to correct the above result by also adding the contribution
of the so-called hard thermal loops \cite{BraPi}.

It is straightforward to check that, in the limit $|\vec{q}| \ll |q_{0}|$,
this HDL polarization tensor indicates the existence of the plasmon
collective modes with $q_{0}^{2} \simeq \omega_{p}^{2}$. In the opposite
limit, we reveal Debye screening (with $m_{D}^{2} = 3 \omega_{p}^{2}$) for
the low frequency electric gluon modes, and no static screening for the
magnetic gluon modes (there is, however, dynamic screening due to Landau
damping).

\subsection{``Light" plasmon mode}
\label{light-plasmon}

As we briefly mentioned earlier, the color flavor locked phase of dense
quark matter reveals additional ``light" plasmons that have no analogues
in ordinary metals. In this subsection, we study these new modes at zero
and finite temperature. We restrict ourselves to only the case of the long
wave length limit.

At $T=0$, the expressions for the component functions in Eqs.~(\ref{H-T})
-- (\ref{M-T}) further simplify:
\ba
H(q_{0}) &=& L(q_{0}) = -1 + \frac{1}{9} \int d \epsilon  \Bigg[
\frac{9|\Delta_{0}|^{2}}{E(4 E^{2}-q_{0}^{2})}
+\frac{(2|\Delta_{0}|^{2}-\epsilon^{2}+\tilde{E}E)
(E+\tilde{E})}{\tilde{E}E[(E+\tilde{E})^{2}-q_{0}^{2}]}
\Bigg],
\label{H-T=0} \\
K(q_{0}) &=& \frac{1}{3} \int d \epsilon  \Bigg[
-\frac{5|\Delta_{0}|^{2}}{E(4 E^{2}-q_{0}^{2})}
+\frac{(2|\Delta_{0}|^{2}+\epsilon^{2}-\tilde{E}E)
(E+\tilde{E})}{\tilde{E}E[(E+\tilde{E})^{2}-q_{0}^{2}]}
\Bigg],
\label{K-T=0} \\
M(q_{0}) &=& 0, \label{M-T=0}
\ea
where $|\Delta_{0}|$ is the value of the color superconducting gap at
zero temperature.

The spectrum of massive (both electric and magnetic type) collective
modes, then, is determined by the equation:
\be
q_0^2 + \omega_p^2 H(q_{0}) = q_0^2 + \omega_p^2 L(q_{0})=0.
\label{eq-T=0}
\ee
As is clear, there is at least one solution to this equation which
corresponds to the ordinary plasmon mode, found earlier. We remind that
such a solution appears at $|q_0| \simeq \omega_p \gg |\Delta_{0}|$,
where $H(q_0) = L(q_0) \simeq -1$. Since the imaginary part of $H(q_0) =
L(q_0)$ is small at large $|q_0|$, this plasmon has a narrow width.

It is more interesting to notice, however, that there is another mode with
the energy less than a threshold of producing a pair of quasiparticles,
$|q_{0}| < 2 |\Delta_{0}|$. Since $|\Delta_{0}| \ll \omega_p$, the
corresponding solution to Eq.~(\ref{eq-T=0}) with a sufficient accuracy is
given by a solution to the approximate equation, $L(q_0)\simeq 0$. Using
Mathematica, we get the solution, $q_0 =m_{\Delta} \approx 1.362
|\Delta_{0}|$. The approximate value of the mass of this mode could
also be extracted from the derivative expansion of the effective action,
obtained in Ref.~\cite{CGN}. Using our approach, the corresponding
derivative expansion appears after expanding the component function in
powers of $q_0$: $L(q_0) \approx L(0) + (q_0/|\Delta_{0}|)^{2}
L_{1}(0)+\ldots$. It is straightforward to check that
\ba
L(0) &=& -\frac{21-8\log2}{54}, \\
L_{1}(0) &=& \frac{21+16\log2}{324},
\ea
so that
\be
q_0 \equiv m_{\Delta} \approx
|\Delta_{0}|\sqrt{6\frac{21-8\log2}{21+16\log2}}
\approx 1.70|\Delta_{0}|.
\label{q_D-approx}
\ee
We see that the derivative expansion somewhat overestimates the value of
the mass. Still the result is below the threshold. Notice that value of
the mass obtained in Ref.~\cite{CGN} was claimed to be $q_0 \approx 2.94
|\Delta_{0}|$ which is larger than our result by a factor of
$\sqrt{3}$.  We believe the authors of Ref.~\cite{CGN} made a simple
arithmetic mistake when substituting their expression (61) into Eq.~(62).

\begin{flushleft}
\begin{figure}
\hbox{
\epsfxsize=8.5cm
\epsffile{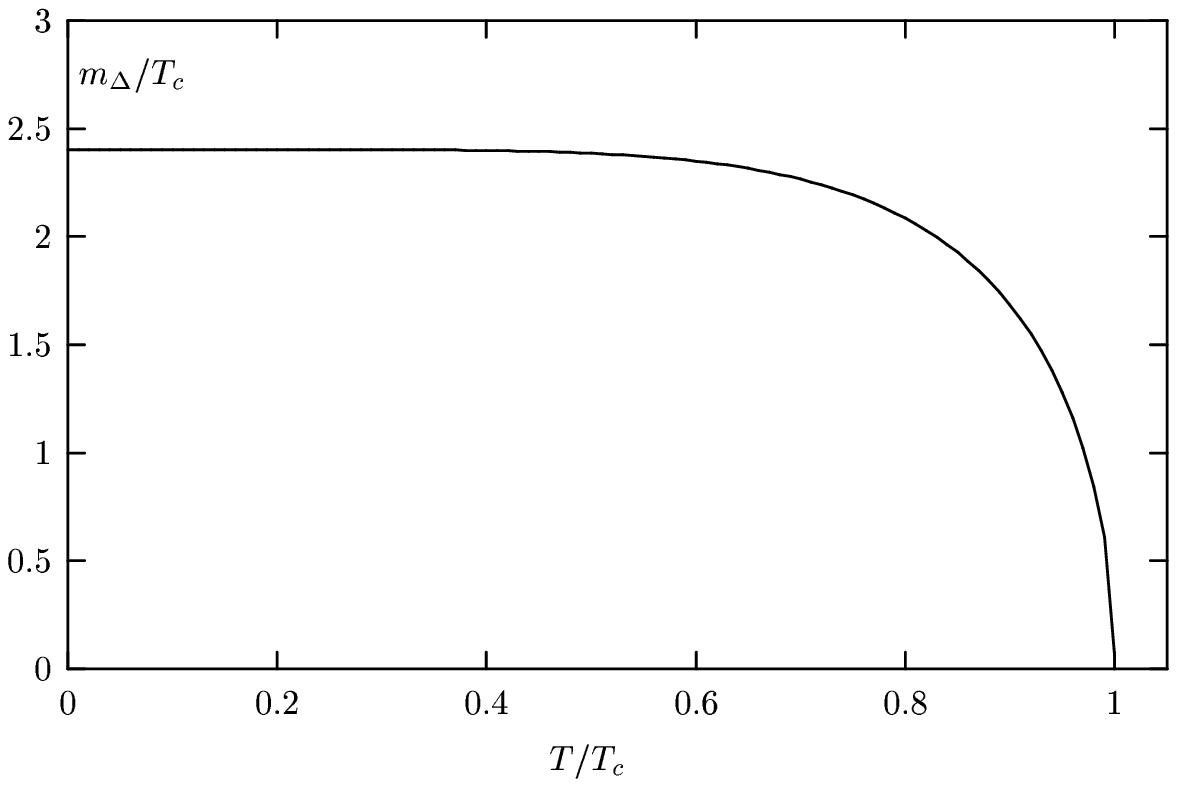}
\quad
\epsfxsize=8.5cm
\epsffile{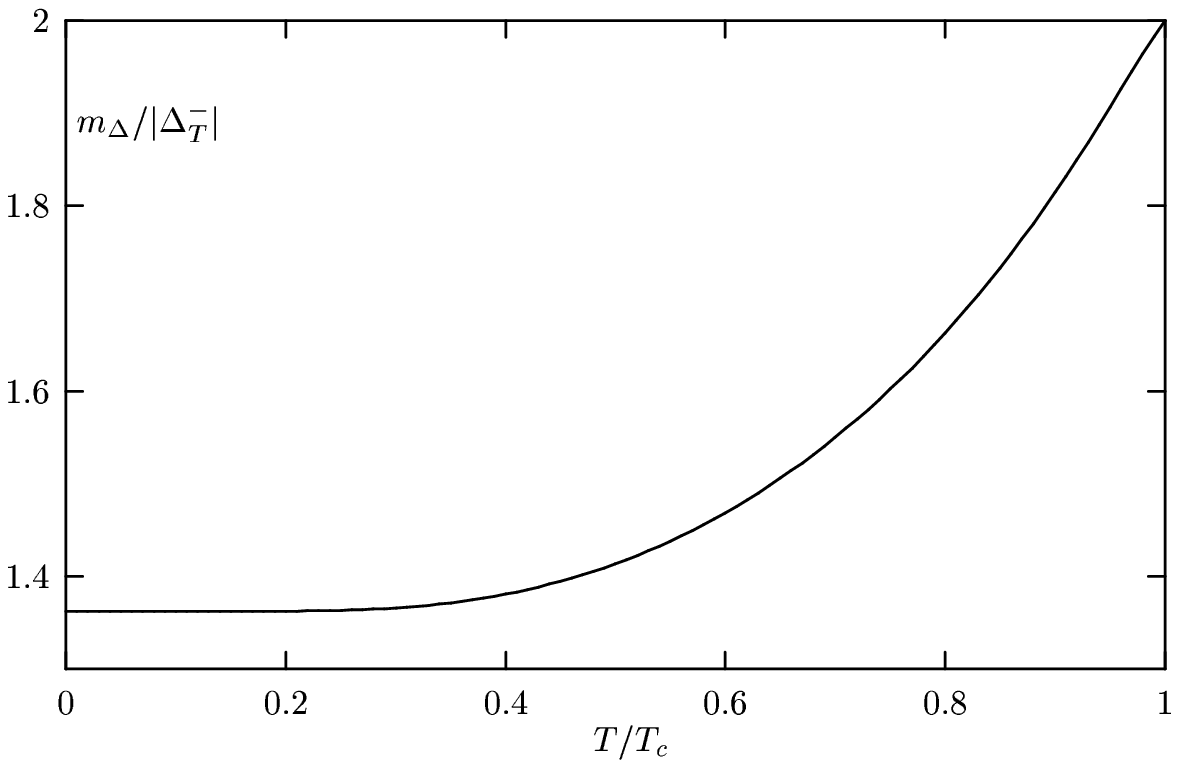} }
\caption{Left: the temperature dependence of the mass of the ``light"
plasmon measured in units of $T_c$. Right: the temperature dependence
of the ratio $m_{\Delta}/|\Delta_{T}|$.}
\label{elm-vs-T}
\end{figure}
\end{flushleft}

By calculating the component functions in Eqs.~(\ref{H-T}) -- (\ref{M-T})
numerically, we also obtained the temperature dependence of the light
massive gluon mode. The result is presented in Fig.~\ref{elm-vs-T}.

\section{NG bosons}
\label{NG-boson}

We recall that the original chiral symmetry is broken in the CFL phase 
of dense QCD. This means that an octet of pseudoscalar NG bosons should
appear in the physical spectrum of our model. Following the approach used
in Sec.~\ref{c-c-function}, we would like to study the axial
current-current correlation function. To this end, we should introduce
axial external sourses $A^{A,\mu}_{5,cl}$ into the microscopic action.
Then, after integrating out the quark degrees of freedom, we would get an
effective action similar to that in Eq.~(\ref{L-g-pi}), except that the
quantum part of the gauge field would be absent. The subsequent
integration over the NG boson field would lead us to a quadratic form
similar to that in Eq.~(\ref{L-g}), determining the axial 
current-current correlation function:
\be
\langle j^{A}_{5,\mu} j^{B}_{5,\nu}
\rangle_{q} = \delta^{AB} \left[ 
\Pi_{1} O^{(1)}_{\mu\nu}(q) 
+ \frac{\Pi_{2} \Pi_{3} + (\Pi_{4})^{2} }{\Pi_{3}}
O^{(2)}_{\mu\nu}(q) \right].
\label{c5c5-function} 
\ee 
The location of the pole in this correlation function defines the
dispersion relation of the NG bosons (which coinsides with the dispersion
relation $q^{\mu} \Pi_{\mu\nu}q^{\nu}=0$, derived in Ref.~\cite{Zar} at
zero temperature),\footnote{We would like to thank
V.A.~Miransky for suggesting this alternative deivation of the NG boson  
dispersion relation.}
\be
q^{\mu}\Pi_{\mu\nu}q^{\nu}
\equiv q^{2}\Pi_{3} \equiv
\omega_{p}^{2}\left[q_{0}^{2} K(q)
- |\vec{q}|^{2} L(q) - 2 q_{0} |\vec{q}| M(q)\right]=0.
\label{NG-gen-disp}
\ee
Here we took into account the representation in Eq.~(\ref{P3-HKLM}). In
the following two subsections, we analytically study this relation at zero
temperature and in the nearcritical region. In the latter case, we also
supplement our analysis by numerical calculations.

\subsection{Nambu-Goldstone modes, $T=0$}
\label{NG-T=0}

At first, let us consider the spectrum of the NG bosons at zero
temperature. In this case, the component function $M(q)$, see
Eq.~(\ref{M-long}) for the definition, is identically equal to zero. As is
clear from Eq.~(\ref{NG-gen-disp}), then, the inverse propagator of the NG
bosons reads
\be
i {\cal D}^{-1}_{NG}(q) = \frac{\omega_{p}^{2}}{g^{2}}
\left[q_{0}^{2} K(q) - |\vec{q}|^{2} L(q)\right].
\ee
In the far infrared region ($|q_{0}|,|\vec{q}| \ll |\Delta_{0}|$), it
is sufficient to use the zero momentum limit of the functions $K(q)$ and
$L(q)$. From Eqs.~(\ref{H-T=0}) and (\ref{K-T=0}), we easily derive these
quantities,
\ba
H(0) &=& L(0) = -\frac{21 - 8 \ln 2}{54},
\label{H0-T=0} \\
K(0) &=& -\frac{21 - 8 \ln 2}{18}.
\label{K0-T=0}
\ea
Finally, by making use of them, we obtain the well known expression
for the decay constant as well as the low-energy dispersion relation of
the NG bosons \cite{SonSt,Zar,bs-cfl}:
\be
F_{NG}^{2} = \frac{\mu^{2}}{2\pi^{2}} \frac{21 -8 \ln 2 }{18},
\quad \mbox{and} \quad
q_{0}^{2} \simeq \frac{1}{3}| \vec{q}|^{2}.
\ee
The above dispersion relation remains almost unchanged even at small
non-zero temperatures. In particular, in Appendix~\ref{AppD}, we show that
the NG boson excitations acquire only an exponentially small width, see
Eq.~(\ref{E13}).\footnote{Notice that the dispersion relation of the NG
bosons (and their width, in particular) might have corrections due to the
interactions of the NG bosons with one another. Since their decay constant
is of the same order as the chemical potential $\mu$, such corrections are
expected to be parametrically suppressed by an inverse power of $\mu$. The
detailed analysis of this point is beyond the scope of this paper.} With
increasing the temperature, this width is expected to grow, reaching its
maximum in the nearcritical region.

\subsection{NG bosons in the nearcritical region}
\label{NG-nearT_c}

The existence of the small parameter $|\Delta_{T}|/T$ in the
nearcritical region is very helpful in applying analytical tools
to studying the properties of collective modes. In the case of the
NG bosons, whose spectrum is determined by
Eq.~(\ref{NG-gen-disp}), we need to calculate the component
functions $K(q)$, $L(q)$ and $M(q)$ as powers expansions in
$|\Delta_{T}|/T$. To this end, it is convenient to utilize the
representation of the component functions in Eqs.~(\ref{K-Is}) --
Eqs.~(\ref{L-Is}). The approximate expressions for $I_{i}$
integrals, with all the corrections up to second order in
$|\Delta_{T}|/T$, are given in Appendix~\ref{AppC}. By making
use of them, we derive the following relation:
\ba
\frac{v_{ng}^2}{3}\left\{-2-\frac{5\pi|\Delta_{T}|}{8T}
+\frac{5\pi|\Delta_{T}|}{8T}\left[1-
\frac{\sqrt{v_{ng}^2-1}}{2v_{ng}}-\frac{v_{ng}}{2}
\arcsin\frac{1}{v_{ng}}\right]
+\frac{49\zeta(3)|\Delta_{T}|^2}{4\pi^2T^2}\right\}+
\frac{7\zeta(3)|\Delta_{T}|^2}{12\pi^2T^2} &\simeq& 0,
\label{goldstone_law}
\ea
where $v_{ng} = q_{0}/|\vec{q}|$ and
$\zeta(3)\approx 1.202$. The corrections to this equation are of
third order in $|\Delta_{T}|/T$. Now, assuming that $v_{ng}$
vanishes as a power of $|\Delta_{T}|$ when $T\to T_c$, we
arrive at a simple quadratic equation that determines the
dispersion relation for the NG bosons,
\be
v_{ng}^{2}+i\frac{5\pi|\Delta_{T}|}{32T} v_{ng}
-\frac{7\zeta(3)|\Delta_{T}|^2}{2\pi^2T^2} =0. \label{disp-NG}
\ee
The solution to this dispersion equation reads
\be
q_{0}
=\frac{|\Delta_{T}|}{T}(\pm x_{ng}^{*} - i y_{ng}^{*})
|\vec{q}|, \label{result-NG}
\ee
where $x_{ng}^{*}\approx 0.215$
and $y_{ng}^{*}\approx 0.245$. Notice that this solution is
consistent with the assumption used in derivation of
Eq.~(\ref{disp-NG}).

\begin{figure}
\hbox{
\epsfxsize=8.5cm
\epsffile{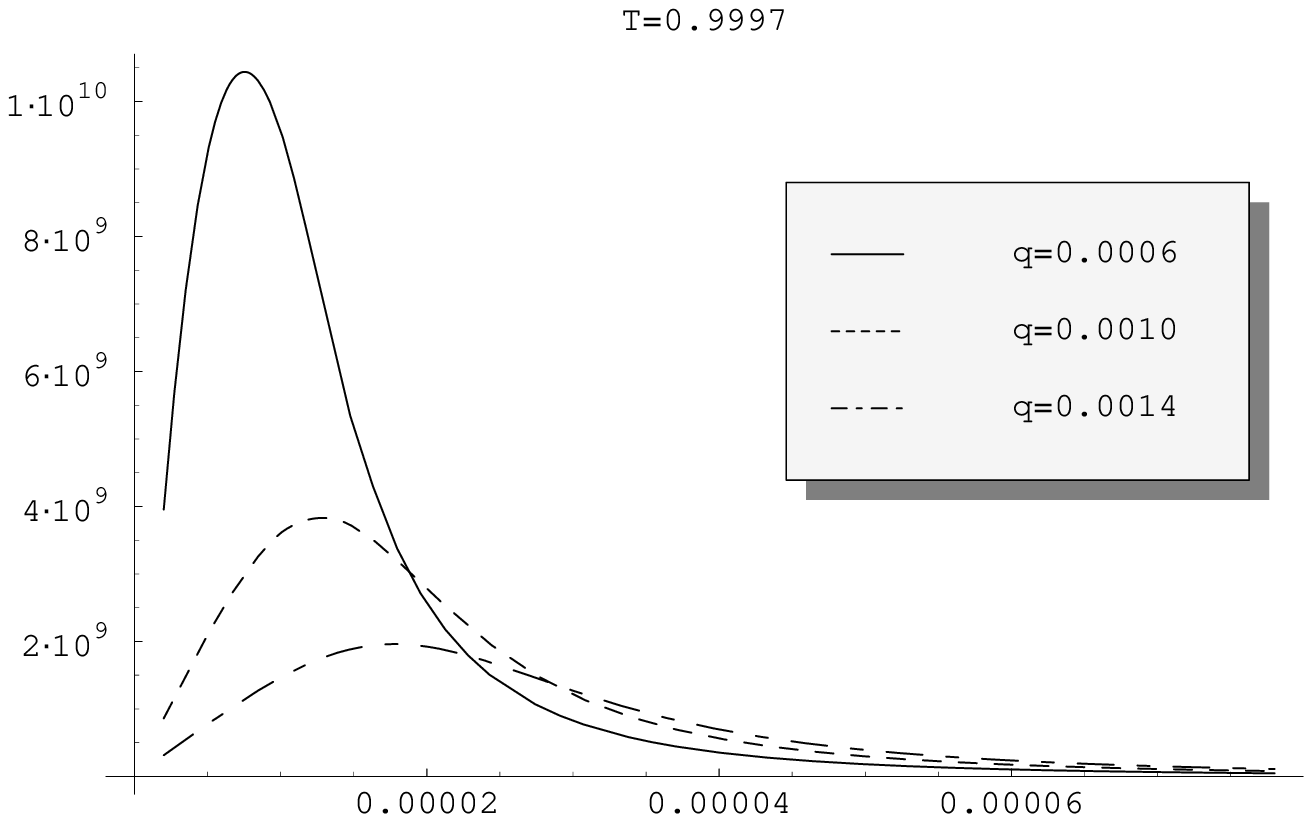}
\quad
\epsfxsize=8.5cm
\epsffile{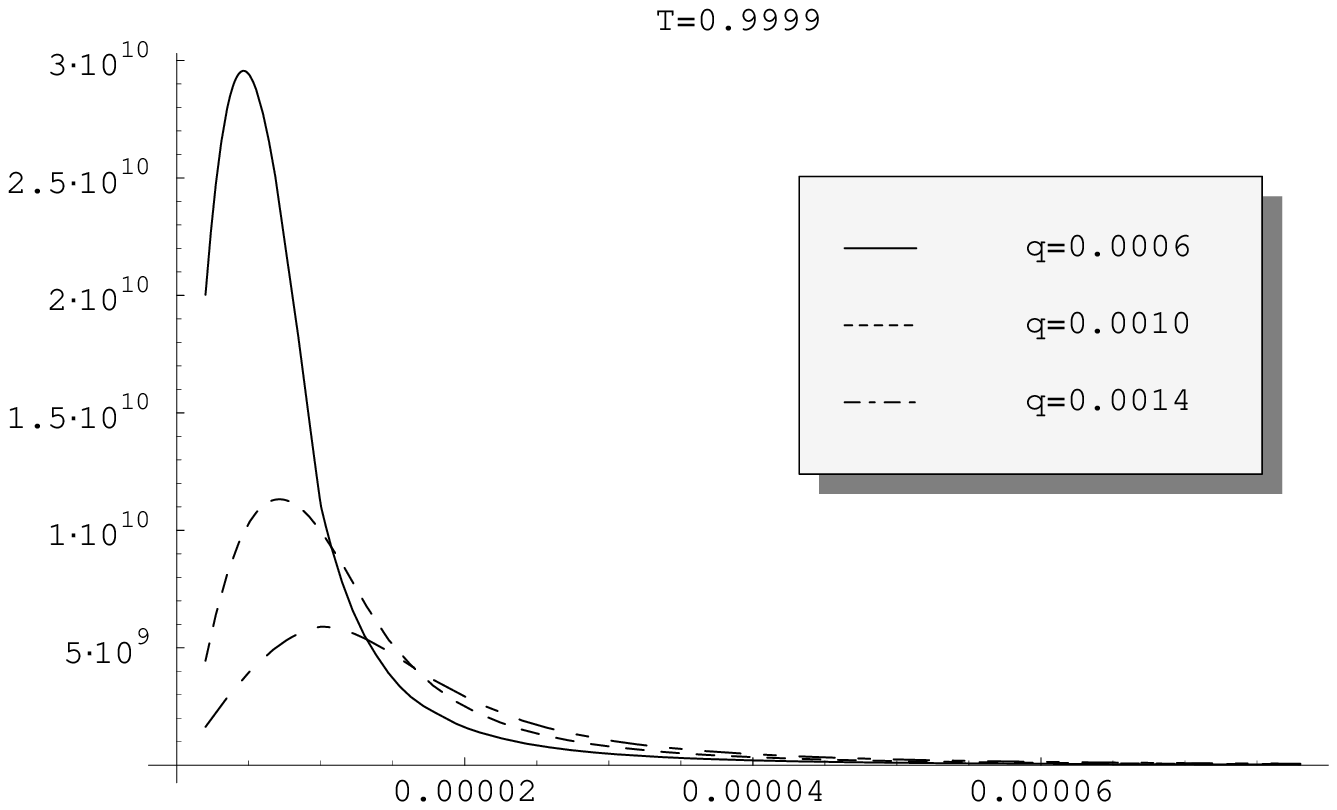} }
\caption{Spectral density of the NG boson at $T=0.9997$ (left) and
at $T=0.9999$ (right) as a function of energy. Everything is
measured in units of $T_{c}$.}
\label{fig-spec-den-NG}
\end{figure}

In order to cross-check our analytical studies, we have also
performed a numerical analysis of the NG boson spectral density.
Such a density is defined by the imaginary part of the
corresponding propagator. In the numerics, we have used the
approximation of small energy and momenta, but we have not used
any assumptions about the value of the superconducting gap. The
results are plotted in Fig.~\ref{fig-spec-den-NG} for two values
of temperature, $T=0.9997 T_{c}$ (corresponding to
$|\Delta_{T}|=0.053 T_{c}$) and $T=0.9999 T_{c}$
(corresponding to $|\Delta_{T}|=0.031 T_{c}$). As is easy to
check, the energy location of the maxima scale with momenta in
accordance with a linear dispersion law that is very similar to
that in Eq.~(\ref{result-NG}).

\section{CG modes}
\label{CG-mode}

Now, let us consider a special type of collective modes, the so-called
Carlson-Goldman gapless modes. Such modes were experimentally discovered
by Carlson and Goldman about a three decades ago \cite{CG} (see also,
Ref.~\cite{Tak97,SchSch,Artem79,Kulik,Tak88,Tak99,Artem}).
One of the most intriguing interpretation connects such modes with a
revival of the Nambu-Goldstone (NG) bosons in the superconducting phase
where the Anderson-Higgs mechanism should commonly take place. It was
argued that an interplay of two effects, screening and Landau damping, is
crucial for the existence of the CG modes \cite{Tak97}. These modes
can only appear in the vicinity of the critical temperature (in the broken
phase), where a large number of thermally excited quasiparticles leads to
partial screening of the Coulomb interaction, and the Anderson-Higgs
mechanism becomes inefficient. At the same time, the quasiparticles induce
Landau damping which usually makes the CG modes overdamped in clean
systems. To suppress such an effect and make the CG modes observable, one
should consider dirty systems in which quasiparticle scatterings on
impurities tend to reduce Landau damping.

In the two fluid description, the CG modes are related to oscillations of
the superfluid and the normal component in opposite directions
\cite{SchSch}. The local charge density remains zero in such oscillations,
providing favorable conditions for gapless modes, in contrary to the
widespread belief that the plasmons are the only collective modes in
charged systems.

The CG modes exist only in the presence of a large number of thermally
excited quasiparticles (normal component). This means, therefore, that
such modes could live only in a finite (possibly very small) vicinity of
the critical temperature.

In the nearcritical region, from Eq.~(\ref{spec-el}) we derive the
following approximate equation for the spectrum of the collective modes of
electric type (assuming that $|\Delta_{T}|/T \ll 1$
and $v_{cg} \ll 1$ where $v_{cg}=|q_{0}|/|\vec{q}|$):
\be
v_{cg}^2+i\frac{45 \pi|\Delta_{T}|}{224T} v_{cg}
-\frac{9\zeta(3)|\Delta_{T}|^2}{8\pi^2T^2} =0.
\label{disp-eq}
\ee
The derivation of this relation is similar to the derivation of
Eq.~(\ref{disp-NG}) in the case of the NG bosons. It might look
surprising, but Eq.~(\ref{disp-eq}) is qualitatively the same as
Eq.~(\ref{disp-NG}). The solution to this dispersion equation
reads
\be
q_{0} =\frac{|\Delta_{T}|}{T}(\pm x_{cg}^{*} - i y_{cg}^{*})
|\vec{q}|,
\label{result}
\ee
where $x_{cg}^{*} \approx 0.193$ and $y_{cg}^{*}\approx 0.316$. This
solution corresponds to the gapless CG modes, and it closely resembles the
dispersion relation of the NG bosons in Eq.~(\ref{result-NG}). The width
of these CG modes is quite large, but this should not be surprising
because we consider the clean limit of the dense quark matter.

\begin{figure}
\hbox{
\epsfxsize=8.5cm
\epsffile{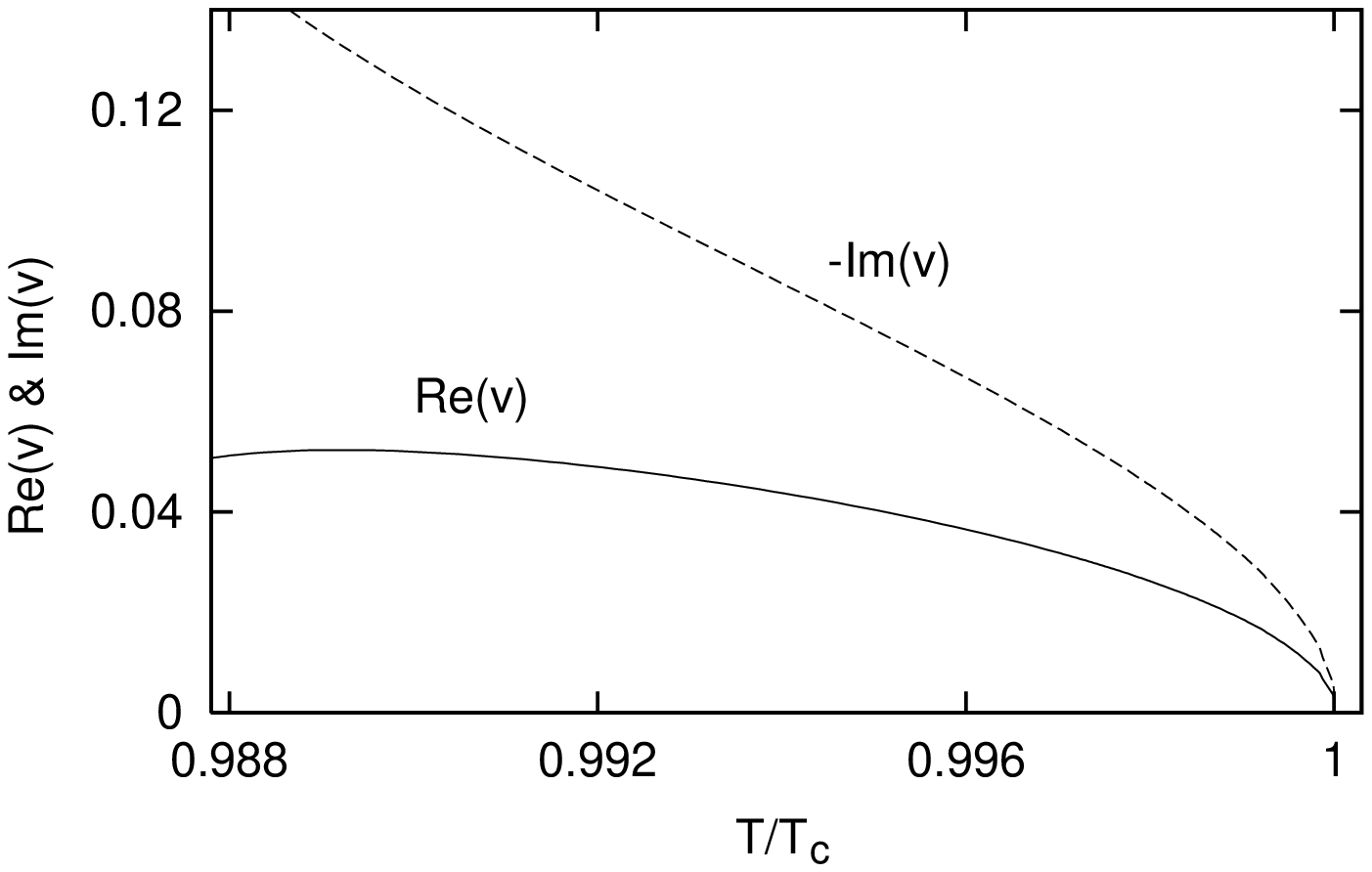}
\quad
\epsfxsize=8.5cm
\epsffile{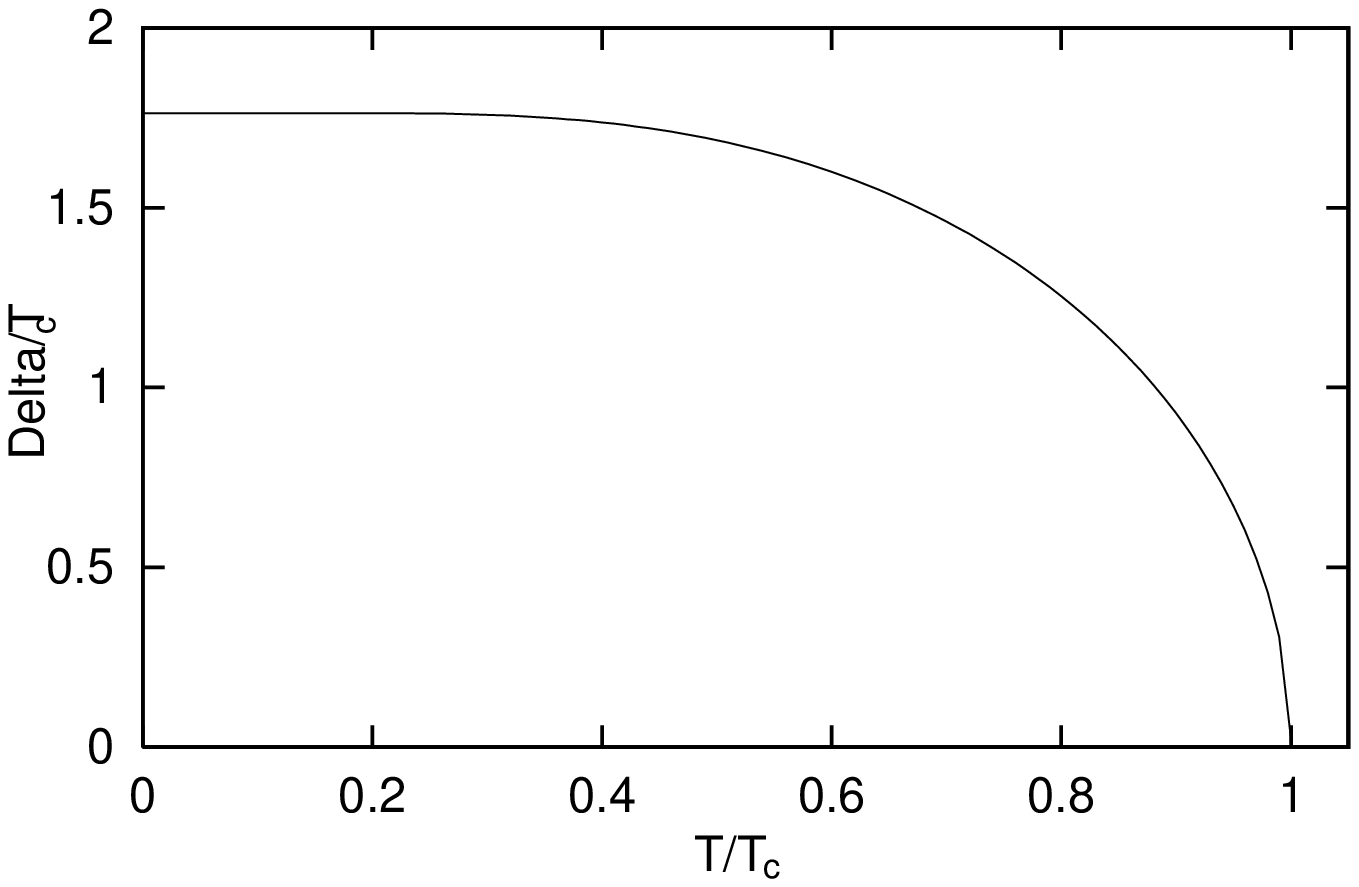} }
\caption{Left: The real (solid line) and imaginary (dash line) parts of
the numerical solution for $v=q_{0}/|\vec{q}|$. Right: Temperature
dependence of the color superconducting gap.}
\label{fig:1}
\end{figure}

The numerical solution for $v_{cg}$ as a function of temperature in
Fig.~\ref{fig:1}. To plot the figure, we used the standard
Bardeen-Cooper-Schrieffer (BCS) dependence of the value of the gap on
temperature, obtained from the following implicit expression:
\be
\ln\frac{\pi T_{c}}{e^{\gamma}|\Delta_{T}|}=\int_{0}^{\infty}
\frac{d\epsilon
\left(1-\tanh\frac{1}{2T}\sqrt{\epsilon^2+|\Delta_{T}|^2}\right)}
{\sqrt{\epsilon^2+|\Delta_{T}|^2}},
\ee
where $\gamma \approx 0.567$ is the Euler constant. As was shown in
Ref.~\cite{PR-weak}, such a dependence remains adequate in the case
of a color superconductor.

It has to be emphasized that the result in Eq.~(\ref{result}) is obtained
for a clean system where Landau damping effects have their full strength.
Therefore, it should have been expected that the CG modes are overdamped.
However, our analysis shows that the presence of two different types of
quarks with nonequal gaps in the CFL phase plays the key role in a partial
suppression of Landau damping. Indeed, as is straightforward to check, 
if the values of the gaps in both the octet and singlet channels were 
equal, the dispersion relation of the CG modes would be given by
\be
v=-i\frac{14\zeta(3)|\Delta_{T}|}{3\pi^{3}T} +\ldots,
\ee
where the higher order terms in powers of $|\Delta_{T}|/T$ are denoted by
the ellipsis. We see from this result that the ratio of the imaginary 
and real parts is much larger than 1 (in fact, it is infinite in this 
leading order approximation), meaning that the CG modes are 
unobservable in the case of equal gaps. In contrast to this, the CG modes
with the dispersion relation in Eq.~(\ref{result}) could be observable
(see our results for the electric gluon spectral density below).

From the numerical result, we clearly see that the ratio of the
width, $\Gamma=-2 Im(v_{cg})|\vec{q}|$, to the energy of the CG
mode, $\varepsilon_{q}=Re(v_{cg})|\vec{q}|$, increases when the
temperature of the system goes further away from the critical
point. At temperature $T^{*} \approx 0.986 T_{c}$, this ratio
formally goes to infinity. This value of $T^{*}$ gives an estimate
of temperature where the CG mode disappears. Of course, such an
estimate is not very reliable since our approximations should
break before the temperature $T^{*}$ is reached. In order to get a
better estimate of temperature where the CG mode disappears, a
more careful numerical calculation is required.

\begin{figure}
\hbox{
\epsfxsize=8.5cm
\epsffile{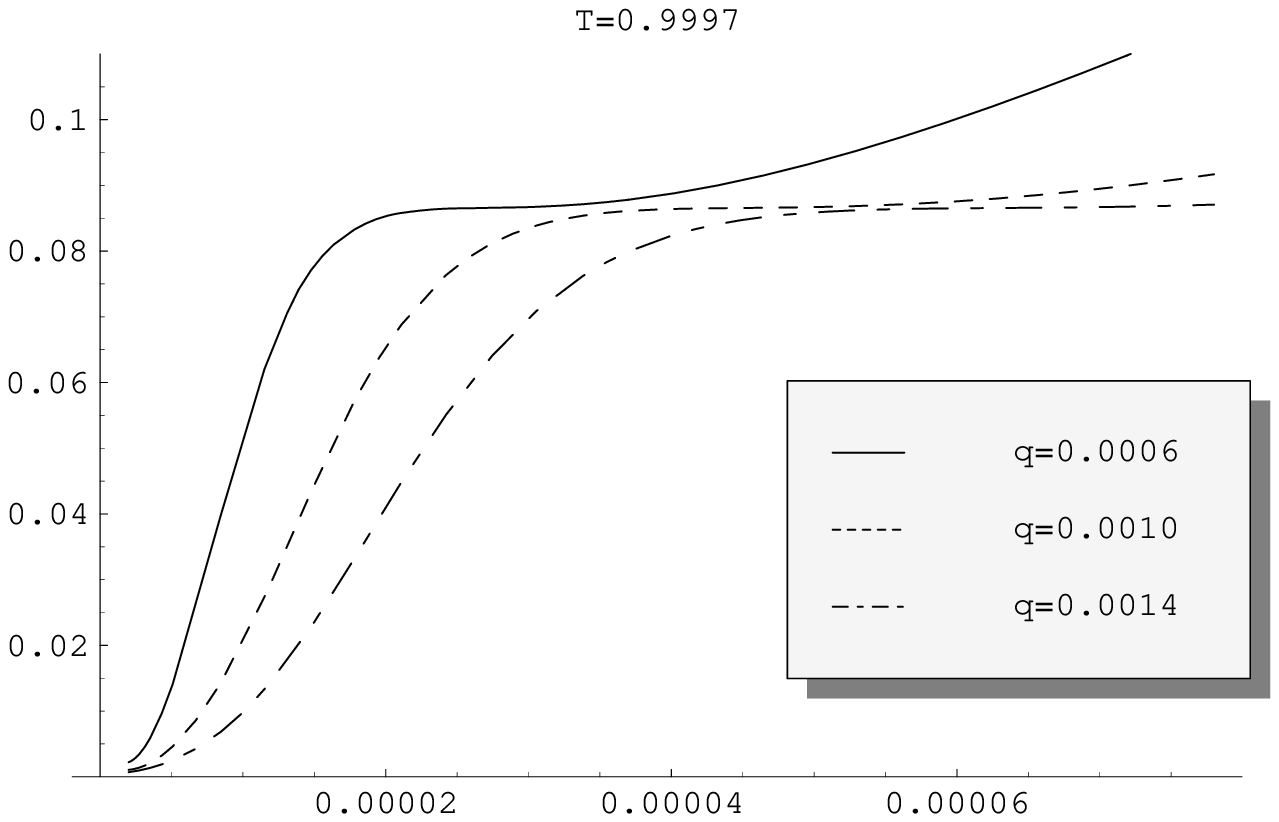}
\quad
\epsfxsize=8.5cm
\epsffile{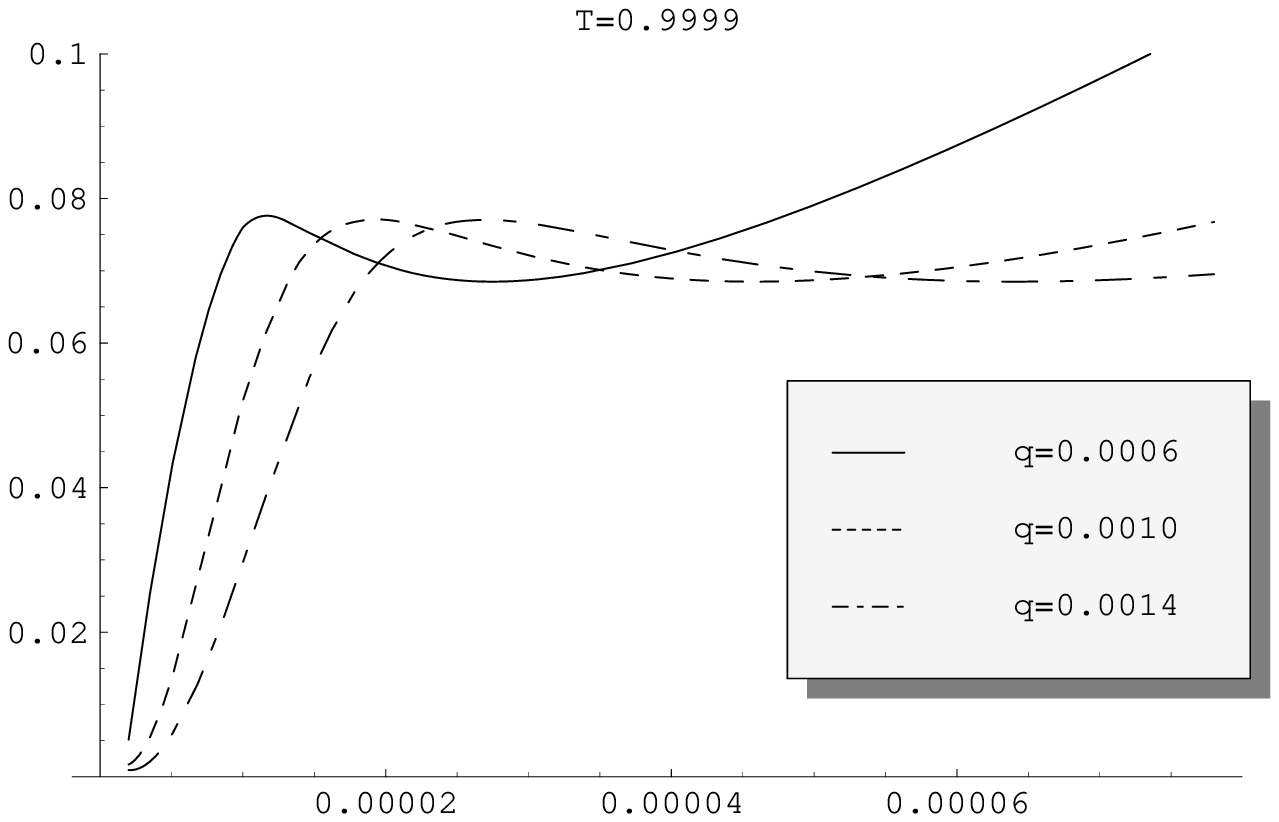} }
\caption{Spectral density of the electric gluon at $T=0.9997$ (left) and
at $T=0.9999$ (right) as a function of energy. Everything is
measured in units of $T_{c}$.}
\label{fig-spec-den-CG}
\end{figure}

As in the case of the NG bosons, we have also performed a numerical
analysis of the electric gluon spectral density in the region of small
momenta where the trace of the CG modes is expected to appear. In our
numerical calculation, we have used the approximation of small energy and
momenta. The results are plotted in Fig.~\ref{fig-spec-den-CG} for two
values of temperature, $T=0.9997 T_{c}$ (corresponding to
$|\Delta_{T}|=0.053 T_{c}$) and $T=0.9999 T_{c}$ (corresponding to
$|\Delta_{T}|=0.031 T_{c}$). Although the maxima in the electron gluon
spectral densities are not so well pronounced as in the case of the
spectral densities of the NG bosons (see Fig.~\ref{fig-spec-den-NG} in the
preceding section), they are detectable when temperatures are sufficiently
close to $T_{c}$.  One could also check that the energy location of the
maxima in Fig.~\ref{fig-spec-den-CG} scale with momenta in accordance with
a linear dispersion law similar to that in Eq.~(\ref{result}).

\section{Meissner effect}
\label{meissner-eff}

Now, let us discuss the Meissner effect at finite temperature. This can be
done by examining the response of the quark system to an external static
magnetic field. In order to derive the expression for the magnetic
$\Pi_{1}(q)$ component of the polarization tensor in the limit $|q_{0}|
\ll |\vec{q}| \to 0$, let us start with the calculation of the
$H(q_0=0,|{\vec q}|)$ function,
\ba
H(|{\vec q}|)&=& -1 + \frac{T}{12} \sum_n
\int \limits_{-1}^{1} d \xi (1-\xi^2) \int d \epsilon
\left[\frac{2|\Delta_{T}|^2-7\epsilon^2+7
\frac{q^2\xi^2}{4}+7\omega_n^2}
{(\omega_n^2+E_{-}^2)(\omega_n^2+E_{+}^2)} \right.
\nonumber\\
&+&\left. \frac{2|\Delta_{T}|^2-\epsilon^2+
\frac{q^2\xi^2}{4}+\omega_n^2}
{(\omega_n^2+E_{-}^2)(\omega_n^2+\tilde{E}_{+}^2)}
+\frac{2|\Delta_{T}|^2-\epsilon^2+
\frac{q^2\xi^2}{4}+\omega_n^2}{(\omega_n^2+\tilde{E}_{-}^2)
(\omega_n^2+E_{+}^2)}\right],
\ea
where $E_{\pm}^2=(\epsilon\pm q\xi/2)^{2}+|\Delta_{T}|^2$ and
$\tilde{E}_{\pm}^2=(\epsilon\pm q\xi/2)^{2}+4|\Delta_{T}|^2$.
The proper calculation of this expression requires performing the
Matsubara summation before integrating over $\epsilon$. This order
could be interchanged if the normal phase contribution
(with $\Delta_{T}=0$) is subtracted from the above expression, i.e.,
\ba
H(|{\vec q}|)&=&H_0(|{\vec q}|)+[H(|{\vec q}|)-H_0(|{\vec q}|)]
=H_0(|{\vec q}|)+\frac{T}{12} \sum_n
\int\limits_{-1}^{1} d \xi (1-\xi^2) \int d \epsilon
\left[\frac{2|\Delta_{T}|^2-7\epsilon^2
+7\frac{q^2\xi^2}{4}+7\omega_n^2}{(\omega_n^2+E_{-}^2)
(\omega_n^2+E_{+}^2)}\right.\nonumber\\
&+&\left.\frac{2|\Delta_{T}|^2-\epsilon^2
+\frac{q^2\xi^2}{4}+\omega_n^2}{(\omega_n^2+E_{-}^2)
(\omega_n^2+\tilde{E}_{+}^2)}
+\frac{2|\Delta_{T}|^2-\epsilon^2
+\frac{q^2\xi^2}{4}+\omega_n^2}{(\omega_n^2+\tilde{E}_{-}^2)
(\omega_n^2+E_{+}^2)}
-9\frac{-\epsilon^2+\frac{q^2\xi^2}{4}+\omega_n^2}
{(\omega_n^2+(\epsilon-q\xi/2)^2)(\omega_n^2
+(\epsilon+q\xi/2)^2)}\right].
\label{H_rearranged}
\ea
It is easy to check that performing summation over Matsubara frequencies
(and taking the limit $|{\vec q}|\to 0$) we come back to
Eq.~(\ref{H-short}) (where we put first $q_0=0$ first and then
$|{\vec q}|=0$). Notice that the last term in square brackets in
Eq.~(\ref{H_rearranged}) reproduces the first term on the right hand
side in Eq.~(\ref{H-short}).

The nice feature about Eq.~(\ref{H_rearranged}) is that it allows to
interchange the order of summation and integration. This is because the
integral and the sum converge rapidly. The initial expression for
$H(q_0,|\vec{q}|)$ did not have this property, and the order of operations
was fixed (first summation, then integration). By evaluating the integral
over $\epsilon$ in Eq.~(\ref{H_rearranged}), we obtain
\ba
H(|{\vec q}|)&=&\frac{\pi
T|\Delta_{T}|^2}{12}\sum_n\int\limits_{-1}^1d\xi(1-\xi^2)
\left\{
-\frac{5/2}{\sqrt{\omega_n^2+|\Delta_{T}|^2}(\omega_n^2
+|\Delta_{T}|^2+q^2\xi^2/4)}\right.\nonumber\\
&+&\left.\frac{6}{9|\Delta_{T}|^4+2q^2\xi^2(2\omega_n^2
+5|\Delta_{T}|^2+q^2\xi^2/2)} \left[\frac{2\omega_n^2
+3|\Delta_{T}|^2+q^2\xi^2/3}{\sqrt{\omega_n^2
+|\Delta_{T}|^2}}-\frac{2(\omega_n^2+
3|\Delta_{T}|^2+q^2\xi^2/3)}{\sqrt{\omega_n^2
+4|\Delta_{T}|^2}}\right]\right\}. \label{energy_int_done}
\ea
From the structure of this expression, it is clear that $|{\vec
q}|/T_c$ is an important dimensionless parameter at all
temperatures. Indeed, for $T \ll T_c$, it is the gap
$|\Delta_{0}|$ which of order $T_c$ that sets the scale in the
$H(|{\vec q}|)$ function. Near $T_c$ (i.e., $|T_{c}-T| \ll
T_{c}$), on the other hand, the gap is small but $\omega_{n\neq
0}=2n\pi T \simeq T_c$ gives the same characteristic scale.

Let us begin by examining the first limiting case, $|{\vec q}|\ll T_c$. In
this limit, we keep only the leading order term on the right hand side of
Eq.~(\ref{energy_int_done}), and drop all terms suppressed by a power
of $|{\vec q}|$. Thus, we arrive at
\ba
H(|{\vec q}|)\simeq\frac{\pi
T|\Delta_{T}|^2}{9}\sum_n\left\{\frac{-5/2}{(\omega_n^2
+|\Delta_{T}|^2)^{3/2}}+
\frac{4}{3|\Delta_{T}|^4}\left(\frac{\omega_n^2
+3|\Delta_{T}|^2/2}{\sqrt{\omega_n^2+|\Delta_{T}|^2}}-
\frac{\omega_n^2+3|\Delta_{T}|^2}{\sqrt{\omega_n^2
+4|\Delta_{T}|^2}}\right)\right\}.
\label{meissner_mass}
\ea
As a benchmark test, let us calculate this expression at $T=0$.
The sum over frequencies in this case can be replaced by an integral,
by setting $2\pi T\delta n=d\omega$. Thus, we arrive at the result
\ba
H(|\vec{q}|\to 0)&=&
\frac{|\Delta_{T}|^2}{18}\int\limits_{-\infty}^\infty
d\omega\left[\frac{-5/2}{(\omega^2+|\Delta_{T}|^2)^{3/2}}
+\frac{4}{3|\Delta_{T}|^4}\left(\frac{\omega^2
+3|\Delta_{T}|^2/2}{\sqrt{\omega^2+|\Delta_{T}|^2}}-
\frac{\omega^2+3|\Delta_{T}|^2}{\sqrt{\omega^2
+4|\Delta_{T}|^2}}\right)\right]\nonumber\\
&=&\frac{1}{9}\int\limits_{0}^\infty dx\left[\frac{-5/2}{(x^2+1)^{3/2}}+
\frac{4}{3}\left(\frac{x^2+3/2}{\sqrt{x^2+1}}-
\frac{x^2+3}{\sqrt{x^2+4}}\right)\right]=-\frac{21-8\log{2}}{54},
\ea
which coincides with the known expression at zero temperature
\cite{SonSt,Zar,bs-cfl,Rsch2}.

Near the critical temperature $T_c$, we expand Eq.~(\ref{meissner_mass})
in powers of $|\Delta_{T}|$ to obtain
\ba
H(|\vec{q}| \to 0)=-\frac{2\pi
T|\Delta_{T}|^2}{3}
\sum_{n=0}^\infty\frac{1}{\omega_n^3}
=-\frac{7\zeta(3)|\Delta_{T}|^2}{12\pi^2T^2}.
\ea
The nonzero value of $H(|\vec{q}| \to 0)$ function in the static limit
indicates that a constant magnetic field is expelled from the bulk of a
superconductor. So, the conventional Meissner effect takes places at
all temperatures $T < T_{c}$ in the color superconducting phase.

For completeness of our presentation, let us also determine the
penetration depth of the magnetic field. To this end, we also need to
calculate the behavior of the $H(|\vec{q}|)$ function for large values of
$|\vec{q}|$. Thus, let us consider the limit $|{\vec q}| \gg T_c$. In this
case, the main contribution to the integral over $\xi$ in
Eq.~(\ref{energy_int_done}) comes from the region $\xi \alt T_c/
|{\vec q}|\ll 1$. In view of this observation, it is justified to neglect
$-\xi^2$ term in the overall factor $(1-\xi^2)$. Then, by
substituting $|{\vec q}|\xi=x$ and expanding the limits of
integration to infinity, we obtain the following approximate
result:
\ba
H(|{\vec q}|)&=&\frac{\pi^2T|\Delta_{T}|^2}{3|{\vec
q}|}\sum_{n\ge0}\left[-\frac{5/2}
{\omega_n^2+|\Delta_{T}|^2}+\frac{1}{\sqrt{\omega_n^2
+|\Delta_{T}|^2}\sqrt{\omega_n^2
+4|\Delta_{T}|^2}} \right. \nonumber \\
&&- \left. \frac{3}{\sqrt{\omega_n^2
+4|\Delta_{T}|^2}\left(\sqrt{\omega_n^2+4|\Delta_{T}|^2}
+\sqrt{\omega_n^2+|\Delta_{T}|^2}\right)}\right],
\label{h-large-q}
\ea
or, written in slightly different form,
\ba
H(|{\vec q}|)&=&\frac{\pi^2|\Delta_{T}|^2}{6|{\vec q}|}\left[-\frac{5}
{4|\Delta_{T}|}\tanh\frac{|\Delta_{T}|}{2T}+T\sum_{n\ge0}
\frac{1}{\sqrt{\omega_n^2+|\Delta_{T}|^2}
\sqrt{\omega_n^2+4|\Delta_{T}|^2}} \right. \nonumber \\
& &\times \left. \left(-1+\frac{9|\Delta_{T}|^2}
{\left(\sqrt{\omega_n^2+4|\Delta_{T}|^2}
+\sqrt{\omega_n^2+|\Delta_{T}|^2}\right)^2}\right)\right].
\label{int_over_x_taken}
\ea
Here we explicitly performed the summation in first term and slightly
rearranged the second term. In calculation leading to
Eq.~(\ref{h-large-q}), we used the following table integrals:
\ba
\int\limits_{-\infty}^\infty\frac{dx}{x^4+2\beta
x^2+1}=\int\limits_{-\infty}^\infty\frac{dxx^2}{x^4+2\beta x^2+1}=
\frac{\pi}{\sqrt{2}}\frac{1}{\sqrt{\beta+1}}.
\ea
At zero temperature the expression (\ref{int_over_x_taken}) reduces to
\ba
H(|{\vec q}|) &=& \frac{\pi |\Delta_{T}|}{6|{\vec
q}|}\left[-\frac{5\pi}{4}
+\int\limits_0^\infty\frac{dx(2+x^2-\sqrt{x^2+4}\sqrt{x^2+1})}
{\sqrt{x^2+4}\sqrt{x^2+1}}\right] \nonumber \\
&=& -\frac{\pi|\Delta_{T}|}{6|{\vec
q}|}\left[\frac{5\pi}{4}+ 2E\left(\frac{\sqrt{3}}{2}\right)
-K\left(\frac{\sqrt{3}}{2}\right)\right].
\ea
Near $T_c$, we expand the expression in Eq.~(\ref{int_over_x_taken}) in
powers of $|\Delta_{T}|$, and obtain
\ba
H(|{\vec q}|)=-\frac{\pi^2|\Delta_{T}|^2}{8|{\vec q}|T}.
\ea
Now, we consider the calculation of the magnetic penetration length.
By making use of the standard definition (see, for example,
Ref.~\cite{Lan-Lif}), we obtain the following estimate:
\ba
\delta=\frac{1}{\pi}\int\limits_{-\infty}^\infty
\frac{dq}{q^2-\omega_p^2H(q)}
=\frac{2}{\pi}\left[\int\limits_{0}^{1/\xi_0}
\frac{dq}{q^2+m_M^2}+\int\limits_{1/\xi_0}^\infty
\frac{dq}{q^2+M^2|\Delta_{T}|/q}\right],
\label{delta}
\ea
where we split the range of integration into two qualitatively different
regions with momenta smaller and larger than the coherence length
$\xi_0\sim 1/T_c$, respectively. The parameters $m_M^2$ and $M^2$ are
defined as follows:
\ba
m_M^2&=& \left\{ \begin{array}{lcc}
\omega_p^2\frac{21-8\log2}{54}, &\mbox{for} & T\ll  T_c, \\
\frac{7\zeta(3)|\Delta_{T}|^2}{12\pi^2T^2}\omega_p^2 , &\mbox{for} & 
T\sim T_c,
\end{array} \right.
\\
M^2 &=& \left\{ \begin{array}{lcc}
M_0^2\equiv \frac{\pi}{6}\left(\frac{5\pi}{4}
+2E(\sqrt{3}/2)-K(\sqrt{3}/2)\right)\omega_p^2,
 &\mbox{for} & T\ll  T_c, \\
\frac{\pi|\Delta_{T}|}{8T}\omega_p^2,  &\mbox{for} & 
T\sim T_c .
\end{array} \right.
\ea
After performing the integration in Eq.~(\ref{delta}), we derive
\ba
\delta &=&  \left\{ \begin{array}{lcc}
m_M^{-1}+\frac{4}{3\sqrt{3}}(M^2|\Delta_{T}|)^{-1/3},
&\mbox{for} & (M^2|\Delta_{T}|)^{1/3}\gg 1/\xi_0 , \\
m_M^{-1}, &\mbox{for} & (M^2|\Delta_{T}|)^{1/3}\ll 1/\xi_0.
\end{array} \right.
\label{delta_length} 
\ea 
At low temperatures, $T\ll T_c$, we have
$1/\xi_0\sim |\Delta_{T}|$ and $m_M^{-1}\sim
M_0^{-1}=\delta_L$ where $\delta_L$ is the London penetration
depth. By taking into account that $M_0 \simeq \omega_{p} \gg
|\Delta_{T}|$, we arrive at the following low temperature
expression for the penetration depth of a magnetic field: 
\ba
\delta &\equiv & \delta_P =
\frac{4}{3\sqrt{3}}(M_0^2|\Delta_{T}|)^{-1/3}, \quad T \ll
T_{c}. 
\ea 
This is the so-called Pippard expression for the
penetration depth (note that $\delta_P\ll \xi_0$). It is easy to
check from the general expression in Eq.~(\ref{delta_length}) that
$\delta=\delta_P$ almost in the whole region of temperatures
$T<T_{c}$. It is also straightforward to check that the London
limit (by definition, $\delta=\delta_L \gg \xi_0$) is realized
only in a very close vicinity of the critical line where
$|T_c-T|\ll T_c$.

\section{Conclusions}
\label{conclusion}

In this paper, we studied collective modes coupled to (vector and axial)
color current in the color-flavor locked phase of cold dense QCD at zero
and finite temperatures. In this class of collective modes, we revealed
and studied the ordinary plasmons, the new type of ``light" plasmons, the
NG bosons and, finally, the gapless CG modes.

As we argued in the main text, the properties of the ordinary (high
frequency) plasmon collective excitations are similar to those in ordinary
metals. The plasma frequency $\omega_{p}$ is proportional to the value of
the chemical potential, and it is essentially independent of the values of
the temperature and/or the superconducting gap in the dense quark matter.
The ``light" plasmon, on the other hand, has very different properties. It
is also a massive excitation (that is why we also call it a plasmon), but
the value of its mass $m_{\Delta}$ is of the order of the superconducting
gap (more precisely, $1.362 |\Delta_{T}| < m_{\Delta} < 2
|\Delta_{T}|$) which is quite small compared to $\omega_{p}$. The
appearance of such a mode seems to be directly related to the existence of
two different gaps in the quasiparticle spectra (in the octet and singlet
channels).

In the long wave length limit $|\vec{q}| \to 0$, the ``light" plasmons are
stable with respect to the decays into the quark type quasiparticles. It
is only because of other possible decay channels (for example, involving
NG bosons), that their width might be nonzero. Now, while we have not
studied the detailed dispersion relation of the ``light" plasmon modes in
the short wave length limit $|\vec{q}| \agt |\Delta_{T}|$, it is
expected that their energy is a monotonically increasing function of
momentum $|\vec{q}|$. If this is true, there should exist a critical value
of the wave length of order $1/|\Delta_{T}|$ at which the energy of
the ``light" plasmon becomes equal to the threshold of the quasiparticle
pair production. The excitations with the wave lengths shorter than the
critical value (and the energy larger than $2|\Delta_{T}|$) could
easily decay into quasiparticles and, therefore, should have a relatively
large width. It means that these new type plasmon modes have two
characteristic plasma frequencies, $m_{\Delta}$ and $2|\Delta_{T}|$.
The stable (narrow width) modes live only in the following window of the
energies, $m_{\Delta} < q_{0} < 2|\Delta_{T}|$.

The properties of the NG bosons in the CFL phase at zero temperature are
well studied in the literature. Here we generalized such studies to the
case of finite temperatures. In particular, we showed that the NG boson
properties at small temperatures remain almost the same as at zero
temperature, having an exponentially small width. With increasing the
temperature, the width also increases. We presented analytical study of
the dispersion relation of NG bosons in the nearcritical region of
temperatures. Our result confirms the general expectation of the slowdown
of the NG bosons in this region, where their maximum velocity is
proportional to the small value of the color superconducting gap.

By making use of an explicitly gauge covariant approach, we also studied
the properties of the gapless CG modes in the CFL phase of cold dense
quark matter in the nearcritical region (just below $T_{c}$) where a
considerable density of thermally excited quasiparticles is present. It is
important to mention that the presence of the CG modes coexists with the
usual Meissner effect, i.e., an external static magnetic field is expelled
from the bulk of a superconductor. The existence of the Meissner effect is
a clear signature that the system remains in the symmetry broken
(superconducting) phase.

In the case of the CFL phase, as we showed, the CG modes appear even in
the clean limit. Despite the sizable width, their traces can be observed
in the spectral density of the electric gluons.  Taking the effect of
impurities into account should, in general, make the CG modes more
pronounced \cite{Tak99}. In realistic systems such as compact stars,
natural impurities of different nature could further improve the
quality of the gapless CG modes.

The existence of a gapless scalar CG modes (in addition to the
pseudoscalar NG bosons) is a very important property of the color
superconducting phase. They may affect thermodynamical as well as
transport properties of the system in the nearcritical region.  In its
turn, this might have a profound effect on the evolution of forming
compact stars. The CG modes might also have a consequence on a possible
existence of the hypothetical quark-hadron continuity, suggested in
Ref.~\cite{SW-cont}. Indeed, one should notice that, in the hadron phase,
there does not seem to exist any low energy excitations with the quantum
numbers matching those of the CG modes.

In the future, it would be interesting to generalize our analysis to the
so-called $S2C$ phase of dense QCD. In absence of true NG bosons in the
$S2C$ phase, the gapless CG modes might play a more important role. Our
general observations suggest that Landau damping should have stronger
influence in the case of two flavors. At the same time, the presence of
massless quarks could lead to widening the range of temperatures where the
CG modes exist. To make a more specific prediction, one should study the
problem in detail.

\begin{acknowledgments}
We would like to thank the members of Physics Department at Nagoya
University, especially Prof. K.~Yamawaki, for their hospitality
during the initial stage of this project. V.P.G. is grateful to
Dr. S. Sharapov for fruitful discussions on CG modes in standard
superconductors.  I.A.S. would like to thank Prof.~A.~Goldman and
Prof.~V.~Miransky for interesting discussions and comments.  This
work was partially supported by Grant-in-Aid of Japan Society for
the Promotion of Science (JSPS) \#11695030. The work of V.P.G. was
also supported by the grants: SCOPES projects 7~IP~062607 and
7UKPJ062150.00/1 of the Swiss NSF and the grant No. PHY-0122450 of
NSF (USA). He wishes to acknowledge JSPS for financial support.
The work of I.A.S. was supported by the U.S. Department of Energy
Grant No.~DE-FG02-87ER40328.
\end{acknowledgments}

\appendix

\section{Identities, formulas, etc.}
\label{AppA}

In the main text we use the following identities for the structure
constants of $SU(3)$:
\ba
d^{ACD} d^{BCD} = \frac{5}{3} \delta^{AB}, \label{iden1} \\
f^{ACD} f^{BCD} = 3 \delta^{AB}. \label{iden2}
\ea
For completeness, we also present the following more complicated
identities:
\ba
d^{AA^{\prime} B^{\prime}} d^{BB^{\prime} C^{\prime}}
d^{CC^{\prime} A^{\prime}} &=& -\frac{1}{2} d^{ABC} ,\\
d^{AA^{\prime} B^{\prime}} d^{BB^{\prime} C^{\prime}}
f^{CC^{\prime} A^{\prime}} &=& -\frac{5}{6} f^{ABC} ,\\
d^{AA^{\prime} B^{\prime}} f^{BB^{\prime} C^{\prime}}
f^{CC^{\prime} A^{\prime}} &=& -\frac{3}{2} d^{ABC} ,\\
f^{AA^{\prime} B^{\prime}} f^{BB^{\prime} C^{\prime}}
f^{CC^{\prime} A^{\prime}} &=& \frac{3}{2} f^{ABC} .
\ea

In this Appendix, we also derive the three types of Matsubara summation
formulas that appear in the main text. By convention, we use the notation
$\omega_{n}=T \pi (2n+1)$, $\Omega_{m}= 2 T \pi m$ and $\beta=1/T$.
The mentioned three types of sums are
\ba
F_{1}(i\Omega_{m};a,b)&=&
T\sum_{n}\frac{1}{\left(\omega_{n}^{2}+a^2\right)
\left[(\omega_{n}+\Omega_{m})^{2}+b^2\right]},
\label{def-F1}\\
F_{2}(i\Omega_{m};a,b)&=&
T\sum_{n}\frac{i\omega_{n}}{\left(\omega_{n}^{2}+a^2\right)
\left[(\omega_{n}+\Omega_{m})^{2}+b^2\right]},
\label{def-F2}\\
F_{3}(i\Omega_{m};a,b)&=&
T\sum_{n}\frac{\omega_{n}(\omega_{n}+\Omega_{m})}
{\left(\omega_{n}^{2}+a^2\right)
\left[(\omega_{n}+\Omega_{m})^{2}+b^2\right]}.
\label{def-F3}
\ea
Let us start from the first one. The sum is performed by
going to the complex plane and introducing the contour
integral. The result is
\ba
F_{1}(i\Omega_{m};a,b)
&=& \frac{ b (\Omega_{m}^{2}-a^2+b^2)\tanh(a\beta/2)
+a(\Omega_{m}^{2}+a^2-b^2)\tanh(b\beta/2)}
{2ab\left[(a+b)^{2}+\Omega_{m}^{2}\right]
\left[(a-b)^{2}+\Omega_{m}^{2}\right]} \nonumber \\
&=& \frac{(a+b)\left[1-n(a)-n(b)\right]}
{2ab\left[(a+b)^{2}+\Omega_{m}^{2}\right]}
+\frac{(a-b)\left[n(a)-n(b)\right]}
{2ab\left[(a-b)^{2}+\Omega_{m}^{2}\right]},
\label{F1-Im}
\ea
where
\be
n(a)=\frac{1}{\exp(a\beta)+1}
\ee
is the Fermi distribution function.

To perform the analytical continuation to the real frequencies,
we substitute $i\Omega_{m}=\Omega+i\eta$ where $\eta$ is a
vanishingly small positive constant. In this way, we arrive at
\ba
F_{1}(\Omega;a,b) &=&  -{\cal P.V.}
 \frac{(a+b)\left[1-n(a)-n(b)\right]}
{2ab\left[\Omega^{2}-(a+b)^{2}\right]}
- {\cal P.V.} \frac{(a-b)\left[n(a)-n(b)\right]}
{2ab\left[\Omega^{2}-(a-b)^{2}\right]} \nonumber \\
&&+i\frac{\pi(a+b)}{2ab}\mbox{sgn}(\Omega)
\delta\left(\Omega^{2}-(a+b)^{2}\right)
\left[1-n(a)-n(b)\right]
\nonumber \\
&&+ i\frac{\pi(a-b)}{2ab}\mbox{sgn}(\Omega)
\delta\left(\Omega^{2}-(a-b)^{2}\right)
\left[n(a)-n(b)\right] .
\label{F1-Re}
\ea
Similarly,
\ba
F_{2}(i\Omega_{m};a,b)&=&
\frac{i\Omega_{m}\left[ 2ab \tanh(a\beta/2)
-(a^2+b^2+\Omega_{m}^{2})\tanh(b\beta/2)\right]}
{2b\left[(a+b)^{2}+\Omega_{m}^{2}\right]
\left[(a-b)^{2}+\Omega_{m}^{2}\right]} \nonumber \\
&=&-\frac{i\Omega_{m}\left[1-n(a)-n(b)\right]}
{2b\left[(a+b)^{2}+\Omega_{m}^{2}\right]}
-\frac{i\Omega_{m}\left[n(a)-n(b)\right]}
{2b\left[(a-b)^{2}+\Omega_{m}^{2}\right]},
\label{F2-Im} \\
F_{3}(i\Omega_{m};a,b)&=&
\frac{ a (a^2-b^2+\Omega_{m}^{2})\tanh(a\beta/2)
+b(b^2-a^2+\Omega_{m}^{2})\tanh(b\beta/2)}
{2\left[(a+b)^{2}+\Omega_{m}^{2}\right]
\left[(a-b)^{2}+\Omega_{m}^{2}\right]} \nonumber \\
&=& \frac{(a+b)\left[1-n(a)-n(b)\right]}
{2\left[(a+b)^{2}+\Omega_{m}^{2}\right]}
-\frac{(a-b)\left[n(a)-n(b)\right]}
{2\left[(a-b)^{2}+\Omega_{m}^{2}\right]},
\label{F3-Im}
\ea
and
\ba
F_{2}(\Omega;a,b)&=& {\cal P.V.}
\frac{\Omega\left[1-n(a)-n(b)\right]}
{2b\left[\Omega^{2}-(a+b)^{2}\right]}
+{\cal P.V.}\frac{\Omega\left[n(a)-n(b)\right]}
{2b\left[\Omega^{2}-(a-b)^{2}\right]}
\nonumber \\
&& -i\frac{\pi|\Omega|}{2b}
\delta\left(\Omega^{2}-(a+b)^{2}\right)
\left[1-n(a)-n(b)\right]
\nonumber \\
&& -i\frac{\pi|\Omega|}{2b}
\delta\left(\Omega^{2}-(a-b)^{2}\right)
\left[n(a)-n(b)\right]
\label{F2-Re} \\
F_{3}(\Omega;a,b)&=& -{\cal P.V.}
 \frac{(a+b)\left[1-n(a)-n(b)\right]}
{2\left[\Omega^{2}-(a+b)^{2}\right]}
+{\cal P.V.} \frac{(a-b)\left[n(a)-n(b)\right]}
{2\left[\Omega^{2}-(a-b)^{2}\right]}
\nonumber \\
&& +i\frac{\pi(a+b)}{2}\mbox{sgn}(\Omega)
\delta\left(\Omega^{2}-(a+b)^{2}\right)
\left[1-n(a)-n(b)\right]
\nonumber \\
&& -i\frac{\pi(a-b)}{2}\mbox{sgn}(\Omega)
\delta\left(\Omega^{2}-(a-b)^{2}\right)
\left[n(a)-n(b)\right].
\label{F3-Re}
\ea
The typical expressions for $a$ and $b$ would be given by
\ba
a &=& \sqrt{\Delta_{1}^{2} + (\epsilon-\xi |\vec{q}|/2)^{2}}, \\
b &=& \sqrt{\Delta_{2}^{2} + (\epsilon+\xi |\vec{q}|/2)^{2}} ,
\ea
where $\epsilon = |\vec{p}|-\mu$, $\xi=\cos\theta$ and $\theta$ is the
angle between the spatial vectors $\vec{p}$ and $\vec{q}$.

The $\delta$-functions that appear in the imaginary terms of
all three sums have the singularities at
\ba
\epsilon^{\pm}_{0} &=&
\frac{ |\vec{q}|\xi (\Delta_{2}^{2}-\Delta_{1}^{2}) \pm \Omega
\sqrt{[\Omega^{2}- |\vec{q}|^{2}\xi^{2}-(\Delta_{1}-\Delta_{2})^{2}]
[\Omega^{2}- |\vec{q}|^{2}\xi^{2}-(\Delta_{1}+\Delta_{2})^{2}]}}
{2(\Omega^{2}- |\vec{q}|^{2}\xi^{2})},
\ea
which is real only when
$\Omega^{2}\leq (\Delta_{1}-\Delta_{2})^{2} + |\vec{q}|^{2} \xi^{2}$,
or when
$\Omega^{2} \geq (\Delta_{1}+\Delta_{2})^{2} + |\vec{q}|^{2} \xi^{2}$.
The first region defines the window of energies in which the Landau
damping operates, while the other region corresponds to energies for
which the on-shell pair production of quasiparticles is allowed.

\section{Operators $O^{(i)}$}
\label{AppZ}

In the main text of the paper, we heavily use the following set of  four
tensors:
\ba
O^{(1)}_{\mu\nu}(q)&=& g_{\mu\nu}-u_{\mu} u_{\nu}
+\frac{\vec{q}_{\mu}\vec{q}_{\nu}}{|\vec{q}|^{2}},
\label{def-O1} \\
O^{(2)}_{\mu\nu}(q)&=& u_{\mu} u_{\nu}
-\frac{\vec{q}_{\mu}\vec{q}_{\nu}}{|\vec{q}|^{2}}
-\frac{q_{\mu}q_{\nu}}{q^{2}},
\label{def-O2} \\
O^{(3)}_{\mu\nu}(q)&=& \frac{q_{\mu}q_{\nu}}{q^{2}},
\label{def-O3} \\
O^{(4)}_{\mu \nu}(q) &=&
O^{(2)}_{\mu \lambda} u^{\lambda} \frac{q_{\nu}}{|\vec{q}|}
+\frac{q_{\mu}}{|\vec{q}|}u^{\lambda} O^{(2)}_{\lambda \nu}.
\label{def-O4}
\ea
The first three of them are the same projectors of the magnetic, electric
and unphysical (longitudinal in a 3+1 dimensional sense) modes of gluons
which were used in Ref.~\cite{us}. In addition, here we also introduced
the intervening operator $O^{(4)}_{\mu \nu}(q)$, mixing the electric and
unphysical modes. In the above definitions of $O^{(i)}_{\mu \nu}(q)$, we
use
\ba
u_{\mu} &=& (1,0,0,0),\\
\vec{q}_{\mu} &=& q_{\mu} - (u\cdot q) u_{\mu}.
\ea
By making use of the explicit form of the operators $O^{(i)}$, it is
straightforward to derive the following relations:
\ba
u_{\mu} u_{\nu} &=&
-\frac{|\vec{q}|^{2}}{q^{2}} O^{(2)}_{\mu \nu}(q)
+\frac{q_{0}^{2}}{q^{2}} O^{(3)}_{\mu \nu}(q)
+\frac{q_{0} |\vec{q}|}{q^{2}} O^{(4)}_{\mu \nu}(q) , \\
\frac{u_{\mu} \vec{q}_{\nu} + \vec{q}_{\mu} u_{\nu}}
{|\vec{q}|} &=& \frac{2 q_{0}|\vec{q}|}{q^{2}}
\left[O^{(2)}_{\mu \nu}(q)-O^{(3)}_{\mu \nu}(q)\right]
-\frac{q_{0}^{2}+ |\vec{q}|^{2}}{q^{2}} O^{(4)}_{\mu \nu}(q),
\ea
as well as the following multiplication rules:
\ba
O^{(1)}(q) O^{(1)}(q) &=& O^{(1)}(q) , \\
O^{(2)}(q) O^{(2)}(q) &=& O^{(2)}(q) , \\
O^{(3)}(q) O^{(3)}(q) &=& O^{(3)}(q) , \\
O^{(1)}(q) O^{(2)}(q) &=& O^{(2)}(q)O^{(1)}(q) =0, \\
O^{(1)}(q) O^{(3)}(q) &=& O^{(3)}(q)O^{(1)}(q) =0, \\
O^{(2)}(q) O^{(3)}(q) &=& O^{(3)}(q)O^{(2)}(q) =0, \\
O^{(4)}(q) O^{(4)}(q) &=& -O^{(2)}(q)-O^{(3)}(q), \\
O^{(4)}(q) O^{(1)}(q) &=& O^{(1)}(q) O^{(4)}(q) =0 ,\\
O^{(4)}(q) O^{(2)}(q) &=& O^{(3)}(q) O^{(4)}(q), \\
O^{(4)}(q) O^{(3)}(q) &=& O^{(2)}(q) O^{(4)}(q).
\ea
In these last expressions, proper contractions of the Lorentz
indices are tacitly assumed.

The structure of the polarization tensor as well as the structure of the
gluon propagator in the main text are given in terms of the $O^{(i)}$
operators. It is very useful, therefore, to know how to invert the
corresponding tensors. By making use of the definitions in
Eqs.~(\ref{def-O1}) -- (\ref{def-O4}), we check that the following
statement is true. If a general matrix (in Lorentz indices) allows the
following decomposition in terms of the operators $O^{(i)}$
\be
A = a O^{(1)}(q) + b O^{(2)}(q) + c O^{(3)}(q) + d O^{(4)}(q),
\ee
its inverse matrix is given by
\be
A^{-1} = \frac{1}{a} O^{(1)}(q) + \frac{c}{bc+d^{2}} O^{(2)}(q)
+\frac{b}{bc+d^{2}} O^{(3)}(q) - \frac{d}{bc+d^{2}} O^{(4)}(q).
\ee

\section{Calculation of $\Pi_{\mu\nu}$}
\label{AppB}

In the Appendix, we calculate the expression of $\Pi_{\mu\nu}$ in
different limits. The complete expression for the polarization tensor
contains a sum of the left- and right-handed contributions [only one of
them is given in Eq.~(\ref{Pi-q}) in the main text]:
\be
\Pi_{\mu\nu}^{AB}(q) = \Pi_{L,\mu\nu}^{AB}(q) + \Pi_{R,\mu\nu}^{AB}(q)
\equiv \delta^{AB} \Pi_{\mu\nu}(q).
\ee
The complete one-loop result (in Minkowski space) is given by
\ba
\Pi_{\mu\nu}(\Omega,q) &=& \frac{g^{2}}{12}
\sum_{e,e^{\prime}=\pm} \int\frac{d^{3}p}{(2\pi)^{3}}
\nonumber \\
&\times&\Bigg\{
e e^{\prime} \epsilon^{e}_{p-q/2} \epsilon^{e^{\prime}}_{p+q/2}
 T^{e,e^{\prime}}_{\mu\nu}
\left[ 7 F_{1}(\Omega;E^{e}_{p-q/2},E^{e^{\prime}}_{p+q/2})
+F_{1}(\Omega;E^{e}_{p-q/2},\tilde{E}^{e^{\prime}}_{p+q/2})
+F_{1}(\Omega;\tilde{E}^{e}_{p-q/2},E^{e^{\prime}}_{p+q/2})
\right]\nonumber \\
&& - T^{e,e^{\prime}}_{\mu\nu}
\left[ 7 F_{3}(\Omega;E^{e}_{p-q/2},E^{e^{\prime}}_{p+q/2})
+F_{3}(\Omega;E^{e}_{p-q/2},\tilde{E}^{e^{\prime}}_{p+q/2})
+F_{3}(\Omega;\tilde{E}^{e}_{p-q/2},E^{e^{\prime}}_{p+q/2})
\right]\nonumber \\
&&+ (e \epsilon^{e}_{p-q/2}+e^{\prime}
\epsilon^{e^{\prime}}_{p+q/2}) U^{e,e^{\prime}}_{\mu\nu}
\left[ 7 F_{2}(\Omega;E^{e}_{p-q/2},E^{e^{\prime}}_{p+q/2})
+F_{2}(\Omega;E^{e}_{p-q/2},\tilde{E}^{e^{\prime}}_{p+q/2})
+F_{2}(\Omega;\tilde{E}^{e}_{p-q/2},E^{e^{\prime}}_{p+q/2})
\right]\nonumber \\
&&+ \Omega e \epsilon^{e}_{p-q/2}  U^{e,e^{\prime}}_{\mu\nu}
\left[ 7 F_{1}(\Omega;E^{e}_{p-q/2},E^{e^{\prime}}_{p+q/2})
+F_{1}(\Omega;E^{e}_{p-q/2},\tilde{E}^{e^{\prime}}_{p+q/2})
+F_{1}(\Omega;\tilde{E}^{e}_{p-q/2},E^{e^{\prime}}_{p+q/2})
\right]
\nonumber \\
&&+2|\Delta_{T}|^{2} V^{e,e^{\prime}}_{\mu\nu}
\left[F_{1}(\Omega;E^{e}_{p-q/2},E^{e^{\prime}}_{p+q/2})
+F_{1}(\Omega;E^{e}_{p-q/2},\tilde{E}^{e^{\prime}}_{p+q/2})
+F_{1}(\Omega;\tilde{E}^{e}_{p-q/2},E^{e^{\prime}}_{p+q/2})
\right]
\Bigg\},
\label{Pi-most-general}
\ea
where the tensor structures are
\ba
T^{e,e^{\prime}}_{\mu\nu} &=& \mbox{tr} \left(
\gamma_{\mu} \gamma_{0} \Lambda^{-e}_{p-q/2} \gamma_{\nu}
\gamma_{0} \Lambda^{-e^{\prime}}_{p+q/2}
\right) + \mbox{tr} \left(
\gamma_{\mu} \gamma_{0} \Lambda^{e}_{p-q/2} \gamma_{\nu}
\gamma_{0} \Lambda^{e^{\prime}}_{p+q/2}
\right), \\
U^{e,e^{\prime}}_{\mu\nu} &=& \mbox{tr} \left(
\gamma_{\mu} \gamma_{0} \Lambda^{-e}_{p-q/2} \gamma_{\nu}
\gamma_{0} \Lambda^{-e^{\prime}}_{p+q/2}
\right) - \mbox{tr} \left(
\gamma_{\mu} \gamma_{0} \Lambda^{e}_{p-q/2} \gamma_{\nu}
\gamma_{0} \Lambda^{e^{\prime}}_{p+q/2}
\right), \\
V^{e,e^{\prime}}_{\mu\nu} &=& \mbox{tr} \left(
\gamma_{\mu} \Lambda^{-e}_{p-q/2} \gamma_{\nu}
\Lambda^{e^{\prime}}_{p+q/2}
\right) +  \mbox{tr} \left(
\gamma_{\mu} \Lambda^{e}_{p-q/2} \gamma_{\nu}
\Lambda^{-e^{\prime}}_{p+q/2}
\right),
\ea
and
\ba
E^{e}_{k} &=&
\sqrt{(\epsilon^{e}_{k})^{2}+|\Delta_{T}|^{2}}, \\
\tilde{E}^{e}_{k} &=&
\sqrt{(\epsilon^{e}_{k})^{2}+4|\Delta_{T}|^{2}}, \ea are the
excitation energies in the octet and the singlet channels,
respectively.

The general expression in Eq.~(\ref{Pi-most-general}) can be approximated
by dropping all the subleading terms.
For large chemical potential, we derive the following approximate
expression:
\ba
\Pi_{\mu\nu}(\Omega,q) &=&
\frac{g^{2} \mu^{2}}{2\pi^{2}}
\left(u_{\mu} u_{\nu} -g_{\mu\nu}\right) \nonumber \\
&&+\frac{g^{2} \mu^{2}}{ 12\pi^{2}}
\int_{-1}^{1} d \xi \int d \epsilon
\Bigg\{ 2 |\Delta_{T}|^{2}
\left(u_{\mu} u_{\nu} (1-\xi^{2}) + \xi^{2} g_{\mu\nu}
+\frac{1-3\xi^{2}}{2} O^{(1)}_{\mu\nu}(q)\right)
\nonumber \\
&\times&
\left[ F_{1}(\Omega;E_{-},E_{+})
+ F_{1}(\Omega;E_{-},\tilde{E}_{+})
+ F_{1}(\Omega;\tilde{E}_{-},E_{+})\right]
\nonumber \\
&+& \left(\epsilon^{2}-\xi^{2} |\vec{q}|^{2}/4\right)
\left(u_{\mu}u_{\nu}(1+\xi^{2})-\xi^{2}g_{\mu\nu}
-\frac{1-3\xi^{2}}{2} O^{(1)}_{\mu\nu}(q)\right)
\nonumber \\
&\times&\left[ 7
F_{1}(\Omega;E_{-},E_{+})
+F_{1}(\Omega;E_{-},\tilde{E}_{+})
+F_{1}(\Omega;\tilde{E}_{-},E_{+})\right]
\nonumber \\
&-&\frac{1}{2} \xi^{2} \Omega
(u_{\nu}\vec{q}_{\mu}+u_{\mu}\vec{q}_{\nu})
\left[ 7 F_{1}(\Omega;E_{-},E_{+})
+F_{1}(\Omega;E_{-},\tilde{E}_{+})
+F_{1}(\Omega;\tilde{E}_{-},E_{+})\right]
\nonumber \\
&-&\left(u_{\mu}u_{\nu}(1+\xi^{2})-\xi^{2}g_{\mu\nu}
-\frac{1-3\xi^{2}}{2} O^{(1)}_{\mu\nu}(q)\right)
\nonumber \\
&\times&
\left[
7F_{3}(\Omega;E_{-},E_{+})
+F_{3}(\Omega;E_{-},\tilde{E}_{+})
+F_{3}(\Omega;\tilde{E}_{-},E_{+})
\right]
\nonumber \\
&+& 2 \epsilon \xi
\frac{u_{\nu}\vec{q}_{\mu}+u_{\mu}\vec{q}_{\nu}}{|\vec{q}|}
\left[7 F_{2}(\Omega;E_{-},E_{+})
+F_{2}(\Omega;E_{-},\tilde{E}_{+})
+F_{2}(\Omega;\tilde{E}_{-},E_{+})
\right]
\Bigg\} . 
\label{Pi-after-T}
\ea

\section{Key integrals in the nearcritical region}
\label{AppC}

In this Appendix, we consider the nearcritical asymptotics of all the
integrals entering the definition of the component functions $H(q)$,
$K(q)$, $L(q)$ and $M(q)$, see Eqs~(\ref{H-Is}) -- (\ref{M-Is}). It is
sufficient for our purposes to consider only the limit of small momenta,
$|q_0|,|\vec{q}|\ll |\Delta_{T}|$, keeping the ratio $v=|q_{0}| /
|\vec{q}|$ arbitrary. By expanding the results in powers of the ratio
$|\Delta_{T}|/T$ and keeping the leading order terms (up to the second
power in $|\Delta_{T}|/T$), we arrive at the following expressions:
\ba
I_{1} &=& \int\limits_{-\infty}^\infty d\epsilon
\frac{|\Delta_{T}|^2\tanh(E/{2T})}{4E^3} \simeq
\frac{\pi |\Delta_{T}|}{8T}
-\frac{7\zeta(3)|\Delta_{T}|^2}{8\pi^2T^2}
+O\left(\frac{|\Delta_{T}|^4}{T^4}\right),\\
I_{2} &=& \int\limits_{-\infty}^\infty
d\epsilon\frac{2|\Delta_{T}|^2+\epsilon^2-E\tilde{E}}
{E\tilde{E}(E+\tilde{E})}\left[1-n(E)-n(\tilde{E})\right]
\nonumber \\
& \simeq & \frac{|\Delta_{T}|}{2T}
\left[ K\left(\frac{\sqrt{3}}{2}\right)
-2 E\left(\frac{\sqrt{3}}{2}\right)\right]
+\frac{7\zeta(3)|\Delta_{T}|^2}{8\pi^2T^2}
+O\left(\frac{|\Delta_{T}|^3}{T^3}\right),\\
I_{3} &=& \int\limits_{-\infty}^\infty
d\epsilon\frac{2|\Delta_{T}|^2-\epsilon^2+E\tilde{E}}
{E\tilde{E}(E+\tilde{E})}\left[1-n(E)-n(\tilde{E})\right]
\nonumber \\
& \simeq & \frac{|\Delta_{T}|}{2T}
\left[K\left(\frac{\sqrt{3}}{2}\right)
+2 E\left(\frac{\sqrt{3}}{2}\right)\right]
-\frac{63\zeta(3)|\Delta_{T}|^2}{8\pi^2T^2}
+O\left(\frac{|\Delta_{T}|^3}{T^3}\right), \\
I_{4} &=& \int\limits_{-\infty}^\infty
d\epsilon\frac{2|\Delta_{T}|^2+\epsilon^2+E\tilde{E}}
{E\tilde{E}(\tilde{E}-E)}\left[n(\tilde{E})-n(E)\right]
\nonumber \\
& \simeq & -2 - \frac{|\Delta_{T}|}{2T}
\left[K\left(\frac{\sqrt{3}}{2}\right)
-2 E\left(\frac{\sqrt{3}}{2}\right)\right]
+\frac{7\zeta(3)|\Delta_{T}|^2}{\pi^2T^2}
+O\left(\frac{|\Delta_{T}|^3}{T^3}\right),\\
I_{5} &=& \int\limits_{-\infty}^\infty
d\epsilon\frac{2|\Delta_{T}|^2-\epsilon^2-E\tilde{E}}
{E\tilde{E}(\tilde{E}-E)}\left[n(\tilde{E})-n(E)\right]
\nonumber \\
& \simeq & 2-\frac{|\Delta_{T}|}{2T}
\left[K\left(\frac{\sqrt{3}}{2}\right)
+2E\left(\frac{\sqrt{3}}{2}\right)\right]
+\frac{7\zeta(3)|\Delta_{T}|^2}{\pi^2T^2}
+O\left(\frac{|\Delta_{T}|^3}{T^3}\right),\\
I_{6}(v) &=& \int\limits_{-\infty}^\infty
d\epsilon\frac{(9|\Delta_{T}|^2+14\epsilon^2)n^\prime(E)}{2E^2}
\left(1-\frac{Ev}{2\epsilon}
\ln\frac{Ev+\epsilon}{Ev-\epsilon}\right)
\simeq  7I_8(v)+\frac{5\pi|\Delta_{T}|}{8T}\left[1
-F\left(\frac{1}{2},\frac{1}{2};\frac{3}{2};\frac{1}{v^2}\right)\right]
\nonumber\\
&-&\frac{35\zeta(3)|\Delta_{T}|^2}{4\pi^2T^2}\left[1
-F\left(\frac{1}{2},1;\frac{3}{2};\frac{1}{v^2}\right)\right]
+O\left(\frac{|\Delta_{T}|^3}{T^3}\right), \\
I_{7}(v) &=& \int\limits_{-\infty}^{\infty}
d\epsilon\frac{(5|\Delta_{T}|^2+14\epsilon^2)n^\prime(E)}{2E^2}
\left(1+\frac{3E^2v^2}{\epsilon^2}-\frac{3E^3v^3}
{2\epsilon^3}\ln\frac{Ev+\epsilon}{Ev-\epsilon}\right)
\simeq -7 + 21 v^{2} I_{8}(v) \nonumber \\
&+& \frac{\pi |\Delta_{T}|}{8T}\left[9+
5 F\left(\frac{1}{2},\frac{3}{2};\frac{5}{2};\frac{1}{v^2}
\right)\right]
-\frac{7\zeta(3)|\Delta_{T}|^2}{4\pi^2T^2}\left[
15 v^2 F\left(\frac{1}{2},1;\frac{3}{2};\frac{1}{v^2}\right)
-15v^2 +2\right]
+O\left(\frac{|\Delta_{T}|^3}{T^3}\right),\\
I_{8}(v) &=& \int\limits_{-\infty}^\infty
d\epsilon\left(1-\frac{Ev}{2\epsilon}
\ln\frac{Ev+\epsilon}{Ev-\epsilon}\right)n^\prime(E)
\simeq -1+F\left(\frac{1}{2},1;\frac{3}{2};\frac{1}{v^2}\right)
-\frac{\pi|\Delta_{T}|}{12 T v^2}
F\left(\frac{3}{2},\frac{3}{2};\frac{5}{2};\frac{1}{v^2}\right)
\nonumber\\
&+&\frac{7\zeta(3)|\Delta_{T}|^2}{4\pi^2T^2}\left[\frac{1}{v^2-1}
+2-2F\left(\frac{1}{2},1;\frac{3}{2};\frac{1}{v^2}\right)\right]
+O\left(\frac{|\Delta_{T}|^3}{T^3}\right), \\
I_{9}(v) &=& \int\limits_{-\infty}^\infty
d\epsilon\frac{(5|\Delta_{T}|^2+14\epsilon^2)n^\prime(E)}{2 E^2}
\left[1-\frac{3E^2v^{2}}{2\epsilon^2}-\frac{3Ev}{4\epsilon}
\left(1-\frac{E^2v^2}{\epsilon^2}\right)\ln\frac{Ev+\epsilon}
{Ev-\epsilon}\right] \simeq \frac{21}{2} I_{8}(v)
-\frac{1}{2} I_{7}(v) \nonumber\\
&+&\frac{27\pi|\Delta_{T}|}{16 T}\left[1
-F\left(\frac{1}{2},\frac{1}{2};\frac{3}{2};\frac{1}{v^2}\right)\right]
-\frac{189\zeta(3)|\Delta_{T}|^2}{8\pi^2T^2}\left[1
-F\left(\frac{1}{2},1;\frac{3}{2};\frac{1}{v^2}\right)\right]
+O\left(\frac{|\Delta_{T}|^3}{T^3}\right)\nonumber\\
&=& \frac{27}{10} I_{6}(v) - \frac{1}{2} I_{7}(v) -\frac{42}{5}
I_{8}(v) , \ea where $K(z)$ and $E(z)$ are the complete elliptic
integrals of the first and second kind, $\zeta(z)$ is the Riemann
zeta function, and $F(a,b;c;z)$ is the hypergeometric function.
Notice that
\ba
K\left(\frac{\sqrt{3}}{2}\right) \approx
2.157,\quad E\left(\frac{\sqrt{3}}{2}\right) \approx 1.211,\quad
\zeta(3) \approx 1.202,
\ea
and
\ba
F\left(\frac{1}{2},1;\frac{3}{2};\frac{1}{v^2}\right) &=&
\frac{v}{2}\ln\frac{v+1}{v-1} , \\
F\left(\frac{1}{2},\frac{1}{2};\frac{3}{2};\frac{1}{v^2}\right) &=&
v \arcsin\left(\frac{1}{v}\right) ,\\
F\left(\frac{3}{2},\frac{3}{2};\frac{5}{2};\frac{1}{v^2}\right) &=&
3v^2\left[\frac{v}{\sqrt{v^2-1}}-v\arcsin\left(\frac{1}{v}\right)
\right],\\
F\left(\frac{1}{2},\frac{3}{2};\frac{5}{2};\frac{1}{v^2}\right) &=&
\frac{3}{2}\frac{v}{\sqrt{v^2-1}}
-\frac{1}{2}F\left(\frac{3}{2},\frac{3}{2};\frac{5}{2};\frac{1}{v^2}\right)\nonumber\\
&=&\frac{3}{2}\left(v^3\arcsin\frac{1}{v}-v\sqrt{v^2-1}\right).
\ea

\section{The results for $I_{1}$ through $I_{9}$ around $T=0$}
\label{AppD}

In this Appendix, we consider the asymptotics of the $I_{i}$
integrals (see Appendix~\ref{AppC} for their definitions) at small
temperatures. The approximate expressions for the first five integrals
are straightforward to derive, provided $T\ll |\Delta_{T}| \approx
|\Delta_{0}|$. The results read
\ba
I_1&\simeq&\frac{1}{2}-\frac{\sqrt\pi}{2}
\left[\left(\frac{2T}{|\Delta_{0}|}\right)^{1/2}
-\frac{3}{4}\left(\frac{2T}{|\Delta_{0}|}\right)^{3/2}\right]
e^{-\frac{|\Delta_{0}|}{T}}, \\
I_2&\simeq&-1+\frac{4}{3}\log2+\frac{\sqrt\pi}{24}
\left(\frac{2T}{|\Delta_{0}|}\right)^{3/2}
e^{-\frac{|\Delta_{0}|}{T}}, \\
I_3&\simeq&1+\frac{4}{3}\log2-\frac{\sqrt{\pi}}{3}
\left[2\left(\frac{2T}{|\Delta_{0}|}\right)^{1/2}
-\frac{13}{16}\left(\frac{2T}{|\Delta_{0}|}\right)^{3/2}\right]
e^{-\frac{|\Delta_{0}|}{T}}, \\
I_4&\simeq&-\sqrt\pi\left[2\left(\frac{2T}{|\Delta_{0}|}\right)^{1/2}
+\frac{3}{16}\left(\frac{2T}{|\Delta_{0}|}\right)^{3/2}\right]
e^{-\frac{|\Delta_{0}|}{T}},\\
I_5&\simeq&\frac{9\sqrt\pi}{16}
\left(\frac{2T}{|\Delta_{0}|}\right)^{3/2}
e^{-\frac{|\Delta_{0}|}{T}}.
\ea
The calculation of the other four integrals is more complicated. For
example, we obtain the following representation for the $I_8(v)$ function:
\begin{eqnarray}
I_8(v)&\simeq&-\frac{2|\Delta_{0}|}{T}\int\limits_0^\infty
dx\left[1-\frac{v\sqrt{x^2+1}}{2x}
\ln\left|\frac{v\sqrt{x^2+1}+x}{v\sqrt{x^2+1}-x}\right|\right.
\nonumber\\
&+&\left.\frac{i\pi v\sqrt{x^2+1}}{2x}
\theta\left(x-v\sqrt{x^2+1}\right)\right]
e^{-\frac{|\Delta_{0}|}{T}\sqrt{x^2+1}}.
\end{eqnarray}
Only the imaginary part of $I_8(v)$ can be calculated exactly
\ba
&&\mbox{Im} I_8(v)=-\pi
v\left\{e^{-\frac{|\Delta_{0}|}
{T\sqrt{1-v^2}}}-\frac{|\Delta_{0}|}{2T}
\left[e^{-\frac{|\Delta_{0}|}{T}}
Ei\left(-\frac{|\Delta_{0}|}{T}(\frac{1}{\sqrt{1-v^2}}-1)\right)
\right.\right.\nonumber\\
&&-\left.\left.
e^{\frac{|\Delta_{0}|}{T}}
Ei\left(-\frac{|\Delta_{0}|}{T}(\frac{1}{\sqrt{1-v^2}}+1)\right)
\right]\right\},
\ea
where $Ei(-z)$ is the integral exponential function. Assuming that
\be
\frac{|\Delta_{0}|}{T}\gg 1 \quad \mbox{and}
\quad
\frac{|\Delta_{0}|}{T}
\left(\frac{1}{\sqrt{1-v^2}}-1\right)\gg 1,
\ee
we derive the following asymptotic behavior of the imaginary part:
\ba
\mbox{Im} I_8(v) \simeq
-\frac{\pi}{v}e^{-\frac{|\Delta_{0}|}{T\sqrt{1-v^2}}}.
\ea
Similarly, we calculate the asymptotics for the functions
$\mbox{Im}I_6(v)$ and $\mbox{Im}I_7(v)$:
\begin{eqnarray}
\mbox{Im}I_6(v) &\simeq&-\frac{\pi}{v}
e^{-\frac{|\Delta_{0}|}{T\sqrt{1-v^2}}}
\left[7-\frac{5}{4} (1-v^2)(1+\sqrt{1-v^2})\right],\\
\mbox{Im}I_7(v) &\simeq&-21\pi
v\left[1+\frac{5}{28}\sqrt{1-v^2}(1+\sqrt{1-v^2})\right]
e^{-\frac{|\Delta_{0}|}{T\sqrt{1-v^2}}}.
\end{eqnarray}
Neglecting for a moment exponentially small temperature corrections
to the real parts, we can write down the dispersion relation for
the NG bosons,
\ba
\frac{21-8\log2}{18}v^2-\frac{21-8\log2}{54}
+i\frac{5\pi}{12}v^3\sqrt{1-v^2}
e^{-\frac{|\Delta_{0}|}{T\sqrt{1-v^2}}}=0.
\ea
This gives the following solution:
\ba
q_0=\frac{|{\vec q}|}{\sqrt3}
\left[1-i\frac{5\sqrt{2}\pi}{4(21-8\log2)}
e^{-\sqrt{\frac{3}{2}}\frac{|\Delta_{0}|}{T}}\right].
\label{E13}
\ea
Note that the dispersion relation of the NG bosons has an exponentially
small imaginary part near $T=0$ (compare with the corresponding dispersion
relation for the Anderson-Bogolyubov mode in ordinary superconductors
\cite{Kulik}). Clearly, this qualitative feature of the result would not
change even after (exponentially) small real contributions are added to
$I_{6}(v)$, $I_{7}(v)$ and $I_{8}(v)$.

\end{document}